\documentclass[journal]{IEEEtran}

\usepackage{epsfig}
\usepackage{graphicx}
\usepackage{amsmath}
\usepackage{times}
\usepackage{amssymb}
\usepackage{amsfonts}

\newtheorem{theorem}{Theorem}
\newtheorem{definition}{Definition}
\newtheorem{lemma}{Lemma}
\newtheorem{proposition}{Proposition}

\newtheorem{corollary}{Corollary}

\newtheorem{remark}{Remark}

\newtheorem{conjecture}{Conjecture}
\newtheorem{example}{Example}

\newcommand{\beqn}{\begin{equation}}
\newcommand{\eeqn}{\end{equation}}

\newcommand{\imp}[1]{\mathsf{imp}({#1}) }
\newcommand{\ie}{{\it i.e.}}
\newcommand{\etal}{\textit{et al. }}

\IEEEoverridecommandlockouts

\begin{document}
\title{Network Coding in a Multicast Switch}
\author{MinJi~Kim,~\IEEEmembership{Student Member,~IEEE,}
        Jay~Kumar~Sundararajan,~\IEEEmembership{Student Member,~IEEE,}
        Muriel~M\'edard,~\IEEEmembership{Fellow,~IEEE,}
        Atilla~Eryilmaz,~\IEEEmembership{Member,~IEEE,}
and~Ralf~K\"{o}tter,~\IEEEmembership{Senior Member,~IEEE}
\thanks{Manuscript received; revised. This material is based upon work that was supported by the National Science Foundation under Grant No. CCR-0325496, Air Force Office of Scientific Research Grant No. FA9550-06-1-0155, DAWN UC Santa Cruz S0176938, Stanford University under the Complex Network Infrastructures for Communication and Power, Sponsor Award No. PY-1362, and DARPA ITMANET. The material in this paper was presented in part at the 2007 IEEE Conference on Computer Communications (INFOCOM), Anchorage, USA, May 2007, and in part at the 2007 IEEE International Symposium on Information Theory (ISIT), Nice, France, June 2007.}%
\thanks{M. Kim, J. Sundararajan, and M. M\'edard are with the Department of Electrical Engineering and Computer Science, Massachusetts Institute of Technology, Cambridge, MA 02139, USA (e-mail: \{minjikim, jaykumar, medard\}@mit.edu).}
\thanks{A. Eryilmaz is with the Department of Electrical and Computer Engineering, Ohio State University, Columbus, OH 43210, USA (e-mail: eryilmaz@ece.osu.edu).}%
\thanks{R. K\"{o}tter is with the Institute for Communications Engineering, Technische Universit\"at M\"unchen, D-80333 Munich, Germany (e-mail: ralf.koetter@tum.de).}%
}
\maketitle

\begin{abstract}
\boldmath The problem of serving multicast flows in a crossbar switch is considered. Intra-flow linear network coding is shown to achieve a larger rate region than the case without coding. A traffic pattern is presented which is achievable with coding but requires a switch speedup when coding is not allowed. The rate region with coding can be characterized in a simple graph-theoretic manner, in terms of the stable set polytope of the ``enhanced conflict graph''. No such graph-theoretic characterization is known for the case of fanout splitting without coding.

The minimum speedup needed to achieve 100\% throughput with coding is shown to be upper bounded by the imperfection ratio of the enhanced conflict graph. When applied to $K\times N$ switches with unicasts and broadcasts only, this gives a bound of $\min(\frac{2K-1}{K},\frac{2N}{N+1})$ on the speedup. This shows that speedup, which is usually implemented in hardware, can often be substituted by network coding, which can be done in software.

Computing an offline schedule (using prior knowledge of the flow rates) is reduced to fractional weighted graph coloring. A graph-theoretic online scheduling algorithm (using only queue occupancy information) is also proposed, that stabilizes the queues for all rates within the rate region.

\end{abstract}


\IEEEpeerreviewmaketitle

\section{Introduction}\label{sec:introduction}
\IEEEPARstart{N}{etwork} information flow is a field of information theory which aims to quantify the maximum information flow through a network. The network information flow problem is closely related to the multi-commodity flow problem and has been studied extensively owing to its wide applications in communication networks.

An information network is represented by a directed graph $\mathcal{N}= (V,E)$ where $(i, j)\in E$ if there is a communication link from node $i \in V$ to node $j\in V$. Each link is associated with a capacity, and we assume that the link is error-free as long as the rate is below this capacity. There are two special subsets $S$ and $T$ of $V$. The set $S$ is the set of sources, which generates mutually independent streams of information or \emph{messages}. The set $T$ is the collection of sinks. Each sink node $t\in T$ requires some subset of the information streams from the source nodes. This is called the \emph{multicast requirement}.

The main question in network information flow is -- given a network $\mathcal{N}= (V,E)$ and a multicast requirement, is it possible to satisfy all the sink nodes without violating the capacity constraints? Before the notion of network coding was introduced, researchers focused on answering this question in a \emph{router network}. A router network is a network where each packet that enters a node can only be routed or relayed onto some outgoing link(s). In other words, the intermediate nodes in the network cannot modify the packets that they receive -- they can only forward the packets. However, Ahlswede \etal \cite{ahlswede} introduced the notion of \emph{network coding}, which allows mixing of data at intermediate network nodes. Section \ref{sec:networkcoding} provides a brief overview of network coding.

\begin{figure}[h!]
\begin{center}
\includegraphics[width=0.43\textwidth]{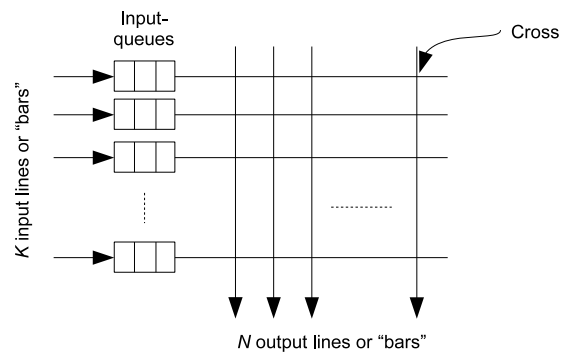}
\end{center}
\caption{A diagram of an $K\times N$ input-queued crossbar switch}\label{fig:crossbar}
\end{figure}

In this paper, we study the benefit obtained from using network coding in a special type of network -- the multicast crossbar switch (see Section \ref{sec:multicastswitchmodel} for background). A crossbar switch is a network of depth one -- it consists of source nodes or the \emph{inputs} and sink nodes or the \emph{outputs}, with every input being directly connected to every output. A crossbar switch with $K$ inputs and $N$ outputs has a $K\times N$ matrix of intersections where the inputs and outputs ``cross'' as shown in Figure \ref{fig:crossbar}. It can be arranged to have what we call the \emph{intrinsic multicast capability} -- an input can convey a packet to several outputs at the same time, by simply connecting the input line to the corresponding output lines. However, an input cannot convey different packets to different outputs at once. The crossbar switch is one of the principal architectures used to construct bigger switches. It is widely used in information processing applications such as telephony and packet switching -- thus, making it an important component of the communication networks, in particular the Internet.

We will focus on input-queued crossbar switches. An input-queued switch is one which has queues at each input to store incoming packets before they are processed by the switching fabric. All input and output lines are assumed to have the same capacity called the line rate. A traffic pattern for which the total rate of flows traversing each input or output is no more than the line rate, is said to be \emph{admissible}. A traffic pattern which can be served without causing the queues to grow unboundedly, is said to be \emph{sustainable} or \emph{achievable}. Note that admissibility is a necessary condition for a traffic pattern to be sustainable. Indeed, if the total rate of flows going into an output exceeds the outgoing line rate, it is physically not possible to keep the queues bounded.

The input-queued crossbar switch has been studied extensively, especially in the context of unicast traffic, where unicast means that for each stream of information, there is only one sink. A unicast traffic pattern is a set of information streams each of which is a unicast flow. It is known that every admissible unicast traffic pattern is also achievable \cite{100forunicast}, \cite{tassephrem}. In other words, as long as no input or output is oversubscribed, the queues can be stabilized, thereby achieving 100\% throughput.

Unfortunately, this result does not extend to multicast flows, where a single stream of information from a source may be destined to reach more than one sink. The extension of the problem from unicast to multicast flows is thus intrinsically more difficult.  Marsan \etal \cite{marsan} showed that 100\% throughput cannot be achieved for multicast flows in an input-queued switch. The authors gave a characterization of the rate region achievable in a multicast switch with \emph{fanout splitting} and also defined the optimal scheduling policy. Fanout splitting is the ability to serve a multicast flow partially to only a subset of its destined outputs, and complete the service in subsequent slots -- see Section \ref{sec:fanoutsplitting} for more details.

Switches have a feature called \emph{speedup} which allows them to process packets faster than the input or output line rate. This feature is usually implemented using parallelization of hardware \cite{speedup}. A formal definition of speedup can be found in Section \ref{sec:speedup}. In \cite{marsan}, the minimum speedup needed to achieve 100\% throughput is shown to grow unboundedly with the switch size for multicast traffic. It is not hard to observe that with enough speedup, a switch can achieve any admissible traffic pattern; however, as it is the case with most hardware features, speedup is expensive to implement and hard to change once the switch is installed.

Another means of increasing throughput in a switch is network coding. We study input-queued switches that are loaded with both unicast and multicast traffic, where inputs are allowed to perform network coding. In this paper, we consider a specific type of network coding -- linear intra-flow network coding for its simplicity and optimality. Note that network coding may be implemented in software, which makes it preferable to speedup as a way to increase the switch throughput. For further details on network coding, see Section \ref{sec:networkcoding}. We ask the question -- what is the magnitude of the benefit we obtain from using network coding in multicast switches? Can we replace speedup with network coding? If not completely, then by how much? Can we use the insight we gain here to design scheduling algorithms for multicast switches with network coding? The main contributions of this paper are:

\begin{enumerate}
\item We prove that linear network coding increases the achievable rate region of a multicast switch. In Section \ref{sec:sim_improverate}, we present an example traffic pattern that demonstrates how network coding increases a switch throughput. In addition, in Section \ref{sec:sim_improvdelay}, we show that network coding allows the switch to be robust to heavy traffic load, resulting in smaller delay compared to fanout-splitting.
\item We propose a graph-theoretic representation of any traffic pattern in terms of what we call the \emph{enhanced conflict graph} and provide a simple graph-theoretic characterization of the multicast switch rate region with coding in Section \ref{sec:networkcodinginswitch}. We prove that the achievable rate region of a network coding multicast switch is a projection of the stable set polytope of the enhanced conflict graph of the traffic pattern.
\item We show that network coding can in many cases substitute for speedup. In Section \ref{sec:codingvsspeedup}, we prove our main result (Theorem \ref{thm:main}) which relates the imperfection ratio of the enhanced conflict graph and the speedup needed to achieve all admissible rates. Using our main result, we provide a lower bound and a graph-theoretic upper bound on the minimum speedup needed to achieve 100\% throughput. In particular, for a $K\times N$ switch with traffic pattern restricted to unicasts and broadcasts only, we show that the minimum speedup is at most $\min \left(\frac{2K-1}{K}, \frac{2N}{N+1}\right)$. This result when applied to a $2\times N$ switch, gives a bound of 1.5 on speedup; however, we conjecture that the actual speedup required to achieve 100\% throughput in a $K\times N$ switch with traffic patterns consisting of unicasts and broadcasts only is 1.25 (Conjecture \ref{thm:5/4} in Section \ref{sec:summaryCodingvsSpeedup}).
\item In Section \ref{sec:algorithms}, we discuss offline and online scheduling algorithms for a multicast switch to achieve the rate region while stabilizing the queues.
\end{enumerate}

As mentioned earlier, for the case of fanout splitting without coding,~\cite{marsan} gave a characterization of the rate region as the convex hull of certain modified departure vectors. However, a graph-theoretic formulation of the same is not known. On the other hand, for the case with coding, our graph-theoretic formulation helps us understand the effect of the traffic pattern on the throughput. The properties of the enhanced conflict graph can be used to derive insight on what kind of traffic patterns are ``difficult'' in terms of computing the schedule, and in terms of achieving $100\%$ throughput.

This paper is organized as follows. Section \ref{sec:background} presents the background and preliminary definitions that will be used in the rest of the paper. This paper mainly draws ideas from network coding (Section \ref{sec:networkcoding}) and graph theory (Section \ref{sec:graphtheory}). Section \ref{sec:networkcodinginswitch} discusses the benefits of network coding when applied to multicast switches. In particular, we present a graph-theoretic formulation of network coding in Sections \ref{sec:cg} and \ref{sec:ecg}. Section \ref{sec:codingvsspeedup} gives the relationship between speedup and imperfection ratio of the enhanced conflict graph, which leads to our main result -- an upper bound on the minimum speedup required to achieve 100\% throughput in a multicast switch with coding. In Section \ref{sec:algorithms}, we use the graph-theoretic formulation of network coding to propose offline and online algorithms for scheduling of a multicast switch. Finally, in Section \ref{sec:conclusion}, we summarize the contributions of this paper and discuss potential avenues for future work.

\section{Preliminaries}\label{sec:background}
This section gives an overview of the relevant work in the area of network coding (Section \ref{sec:networkcoding}), multicast switching theory (Section \ref{sec:multicastswitchmodel} and \ref{sec:multicastswitchscheduling}), and graph theory (Section \ref{sec:graphtheory}).

\subsection{Network coding}\label{sec:networkcoding}

Reference \cite{ahlswede} showed that coding within a network allows a source to multicast information at a rate approaching the smallest cut between the source and any receiver, as the coding field size approaches infinity. Li, Yeung and Cai \cite{LYC} showed that any solvable network with one source and multiple sinks (called \emph{multicast network}) has a scalar linear solution over a sufficiently large finite field alphabet. In addition, \cite{LYC} showed that in multicast networks, linear coding suffices to achieve the optimum, which is the max-flow from the source to each sink. Subsequently, K\"{o}tter and M\'{e}dard \cite{algebraic} showed that in the general network coding problem, deciding achievability and solvability is equivalent to deciding whether a certain algebraic variety is empty or not. Noting the potential of linear network coding, they presented an algebraic framework for linear network coding in arbitrary networks and showed that a simple linear code is sufficient to achieve capacity in the multicast problem.

As a result, there has been a great emphasis on linear network coding. For instance, Ho \etal \cite{rlc} proposed a simple, practical capacity-achieving code. They proposed that every node construct its linear code randomly and independently from all other nodes. This simple construction was shown to achieve capacity with probability exponentially approaching 1 with the field size. M\'{e}dard \etal \cite{conjecture} conjectured that every solvable network has a linear solution over some finite field alphabet and vector dimensions. However, Dougherty \etal \cite{insufficiency} provided a counterexample non-multicast network which is not solvable with linear coding. Although \cite{insufficiency} proved that linear network coding is not sufficient for general networks, linear network coding nevertheless is still a powerful tool.  In particular, if only intra-session coding is allowed, linear network coding suffices for networks with multiple multicast sessions, including multicast switches. Linear intra-session coding for multiple multicast networks was studied in \cite{harish}. In our paper, we only allow intra-flow coding, \ie, packets are coded together only if they have the same source and destination set. Therefore, we shall only consider linear codes.

\subsection{Multicast switch model}\label{sec:multicastswitchmodel}
Multicast switches can be thought of as simple information networks where there are only sources and sinks, no intermediate nodes. Each source is connected to all sinks. In the most basic model, a switch acts as a router. We will now formally specify the switch model used in this paper.

A $K\times N$ switch consists of $K$ sources or \emph{inputs} and $N$ sinks or \emph{outputs}. Packets arrive at inputs on input lines, and depart from outputs on output lines. All input and output lines are assumed to have the same capacity called the \emph{line rate}. We consider a slotted time system, where the length of the slot is chosen to be the reciprocal of the line rate. Henceforth, all rates will be normalized with respect to the line rate, and will expressed in packets per slot. All packets are assumed to be of the same size. The speed of the switch fabric is assumed to be such that if it connects an input to an output, it can transfer one packet over this connection, in one slot. This corresponds to a speedup of 1 (Speedup is defined in Section \ref{sec:multicastswitchscheduling}). Arrivals may occur any time during a slot. All transmissions are assumed to begin just after the beginning of a slot and end just before the end of the same slot. The switch configuration may change only at slot boundaries.

\begin{definition}[\textit{\textbf{Rate}}]
A \emph{rate} specifies the average number of packets that needs to be transferred from an input to the outputs per slot. A rate of 1/2, for example, means that on average the input has to send one packet over two slots.
\end{definition}

\begin{definition}[\textit{\textbf{Flow}}]
A \emph{flow} is the stream of all packets that have a given input and a given destination set. Thus, a flow is specified by a 2-tuple $(i, J)$ consisting of the input $i$ and a set $J$ of outputs corresponding to the destination set of the multicast stream. This set $J$ of outputs is called the \emph{fanout set}. Sometimes, we denote a flow by a 3-tuple, $(r, i, J)$ where $r$ is the rate of the flow. For example, in a $2 \times 3$ switch, we could have a flow $f = (1/2, 1, \{1, 2\})$ which is a stream of packets from input 1 to outputs 1 and 2 with a rate of 1/2.
\end{definition}

\begin{definition}[\textit{\textbf{Subflow}}]
A \emph{subflow} of flow $(i, J)$ is the part of a flow from input $i$ that goes to a particular output $j$ in $J$. Therefore, a subflow is specified by a 3-tuple $(i, J, j)$ consisting of the input $i$, the fanout $J$ and one output $j \in J$. The rate of a subflow is defined to be the rate of the flow to which it belongs. Sometimes, we denote the subflow by a 4-tuple $(r, i, J, j)$, where $r$ is the rate of the subflow. For instance, a flow $f = (1/2, 1, \{1, 2\})$ has two subflows associated with it: $f_1 = (1/2, 1, \{1, 2\}, 1)$ and $f_2 = (1/2, 1, \{1, 2\}, 2)$.
\end{definition}

The constraints on the switch configuration are specified below:
\begin{itemize}
\item An input may send the same packet to many outputs at once, but may not send different packets to different outputs simultaneously. This is called the \emph{intrinsic multicast capability}.
\item An output may receive a packet from only one input at a time.
\end{itemize}

\begin{figure}[t]
\begin{center}
\includegraphics[width=0.45\textwidth]{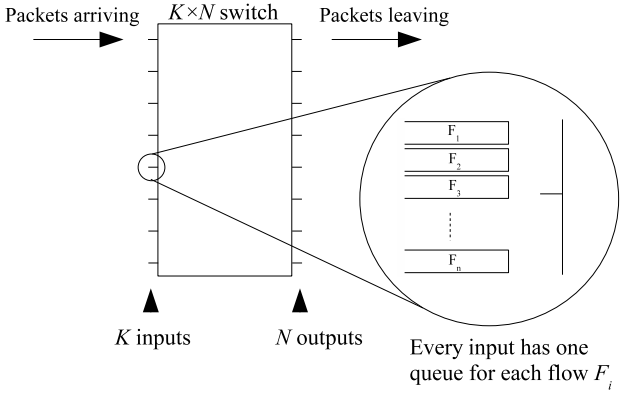}
\end{center}
\caption{$K\times N$ input-queued multicast switch}\label{fig:switch}
\end{figure}

These constraints give rise to the need for queues at the inputs as multiple packets may arrive at an input simultaneously. Each input maintains a separate queue for each flow. Therefore, if we have every possible flow through an input, then the input needs to maintain a set of $2^N-1$ queues; otherwise, fewer queues will suffice. The queues are assumed to have infinite capacity, but the goal of the scheduling algorithms will be to keep their occupancy stable. A diagram of a $K \times N$ input-queued multicast switch is given in Figure \ref{fig:switch}.

\begin{definition}[\textit{\textbf{Traffic Pattern}}]
A \emph{traffic pattern} is a collection of flows. A traffic pattern is called \emph{admissible} if the sum of the rates of all the flows through each input or output does not exceed one, \ie, the inputs and outputs are not oversubscribed. A traffic pattern is said to be \emph{achievable} if there exists a switch schedule that can serve it, while keeping the queues stable.
\end{definition}

\subsection{Scheduling strategies}\label{sec:multicastswitchscheduling}
Clearly, admissibility is a necessary condition for a traffic pattern to be achievable; however, it is not clear whether the converse holds. It turns out that the converse is true for unicast traffic \cite{100forunicast}, but not for multicast traffic \cite{marsan}.

For unicast traffic, Chang \etal \cite{bvnswitches} presented a scheme, called the Birkhoff-von Neumann switch, that not only achieves 100\% throughput but also guarantees packet delay in offline settings. The Birkhoff-von Neumann switch is based on a theorem that says any doubly stochastic matrix can be expressed as a convex combination of permutation matrices \cite{birkhoff}\cite{neumann}. Note that, any admissible unicast traffic pattern can be converted to a doubly stochastic matrix. Then, the doubly stochastic matrix is decomposed into permutation matrices, which in turn correspond to switch states.

Sundararajan \etal \cite{jaykumar} extended this Birkhoff-von Neumann approach to multicast switching. Using a graph-theoretic formulation, they showed that the rate region of multicast switching without fanout splitting (defined in Section \ref{sec:fanoutsplitting}) is precisely the stable set polytope of the traffic pattern's ``conflict graph'', which we shall discuss in Section \ref{sec:cg}. As a result, they showed that the problem of deciding achievability in a multicast switch is equivalent to the membership problem for the stable set polytope of a graph, which is known to be $NP$-hard. In addition, \cite{jaykumar} showed that computing the offline schedule for multicast traffic, unlike that for unicast traffic, is hard. Indeed, it is equivalent to fractional weighted graph coloring, which is $NP$-hard in general. Thus, many of the complexity and achievability results for unicast traffic do not extend to multicast traffic. Even if a traffic pattern is admissible, depending on the switch's capabilities, the switch may not be able to achieve the traffic pattern.

\begin{example}
Consider the traffic pattern shown in Figure \ref{fig:impossible}. This traffic pattern consists of a broadcast flow $(1/2, 1, \{1,2\})$, and two unicast flows $(1/2, 2, \{1\})$ and $(1/2, 2, \{2\})$. This traffic pattern shown in Figure \ref{fig:impossible} is admissible since every input and output has a total rate of at most 1. However, if the switch is restricted to serve the broadcast flow to all outputs at once, \ie, it is not allowed to split the fanout, then at most one of the three flows can be served at a time. In this case, the sum of the rates of the three flows must be less than 1 to be achievable; however, the sum of the rates is 3/2. The broadcast from input 1 at rate 1/2 requires half of the time. During this time, input 2 cannot serve the two unicasts. But that leaves input 2 with the remaining half of the slots to serve two unicasts at rate 1/2 each, which is not possible. Therefore, this traffic pattern is not achievable.

\begin{figure}[h!]
\begin{center}
\includegraphics[width=0.50\textwidth]{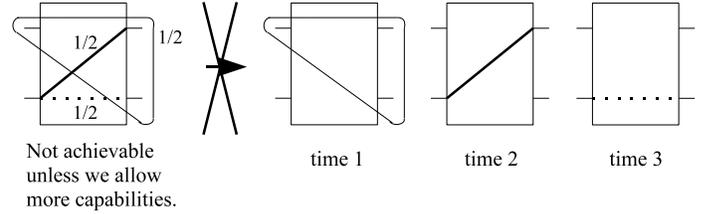}
\end{center}
\caption{Admissible but not-achievable traffic pattern}\label{fig:impossible}
\end{figure}

\end{example}
This observation that not all admissible traffic patterns are achievable raises the question of how much of the admissible rate region is actually achievable. To achieve those admissible but not achievable traffic patterns, what additional capabilities does a switch require? What capability of a switch is the most effective in increasing the achievable rate region to be at least the admissible rate region? In Sections \ref{sec:fanoutsplitting}, \ref{sec:intraflowcoding} and \ref{sec:speedup}, we present three approaches -- fanout splitting, linear network coding, and speedup -- to increase the rate region of a switch.

\begin{figure}[tbp]
\vspace*{0.3cm}
\begin{center}
\includegraphics[width=0.50\textwidth]{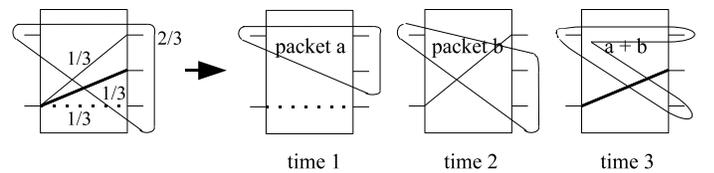}
\end{center}
\caption{A traffic pattern that shows the benefit of coding}
\label{fig:codinggood}
\end{figure}

\subsubsection{Fanout splitting}\label{sec:fanoutsplitting}

There are many ways in which a multicast switch can serve a multicast flow. The most simple method would be to serve all the multicast flow as if it was multiple unicast flows. For example, the packets of $f = (1/2, 1, \{1, 2\})$ could be ``copied'' into two separate unicasts $f_1 = (1/2, 1, \{1\})$ and $f_2 = (1/2, 1, \{2\})$. This scheme is inefficient because, in some cases, it converts an originally achievable traffic pattern into one that is inadmissible. For example, copying $f' = (1/2, 1, \{1, 2, 3\})$ into three unicasts will make three flows with rate 1/2 which overbooks input 1. The other extreme is to force the input to send the multicast packet to every output node in the fanout set simultaneously, which was described as the \emph{no-splitting} strategy in \cite{fanoutsplitting}. However, this scheme can be restricting, as shown by the example in Figure \ref{fig:impossible}.

\begin{figure}[tbp]
\begin{center}
\includegraphics[width=0.45\textwidth]{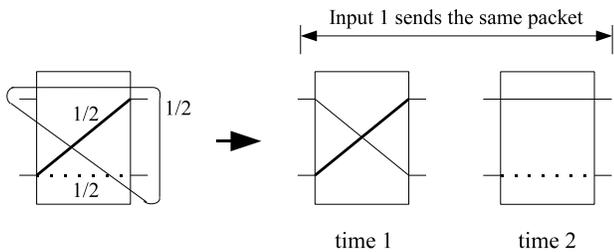}
\end{center}
\caption{A traffic pattern which demonstrates the benefit of
fanout-splitting}\label{fig:fsgood}
\end{figure}

The middle ground between copying and no-splitting is \emph{fanout-splitting} \cite{fanoutsplitting}. Fanout-splitting allows the source to serve subsets of the fanout set at different points in time. Therefore, copying and no-splitting are two extreme cases of fanout-splitting: the first serves the fanout set by dividing it into subsets of size one, the latter serves it by not splitting at all. By definition, fanout-splitting achieves a greater rate region than copying or no-splitting.

\begin{example}
The pattern in Figure \ref{fig:impossible} cannot be satisfied by a no-splitting strategy, but with fanout-splitting this traffic pattern can be achieved as shown in Figure \ref{fig:fsgood}. In Figure \ref{fig:fsgood}, we can see that input 1 completes the broadcast over two slots using fanout-splitting, while input 2 serves unicasts to the idle outputs over the same two slots.
\end{example}

However, even with fanout-splitting, some admissible traffic patterns are not achievable. Figure \ref{fig:codinggood} gives an example of such a case.

\begin{example}
The traffic pattern in Figure \ref{fig:codinggood} is very similar to that in Figure \ref{fig:impossible}, however, with one more output. In order for input 2 to complete all three unicasts, input 2 needs to be serving one of the unicasts at all times. As a result, in each slot, input 1 can partially serve its broadcast packet to at most two idle outputs. Therefore, to serve each broadcast packet completely, input 1 requires two slots. This implies that input 1 can serve the broadcast flow at rate at most 1/2, even if it is allowed to use fanout-splitting; however, the traffic pattern shown in Figure \ref{fig:codinggood} requires a broadcast rate of 2/3.
\end{example}

\subsubsection{Linear network coding}\label{sec:intraflowcoding}

In this paper, we consider a model where the switch, in addition to fanout splitting, is allowed to perform \emph{linear intra-flow coding}, \ie, inputs can now code across packets from the same flow. In the rest of this paper, network coding means linear intra-flow coding. The benefit of network coding can be seen in Figure \ref{fig:codinggood}. It illustrates a schedule that achieves the traffic pattern which we showed cannot be achieved using just fanout-splitting. It is important to note that linear network coding requires fanout-splitting. If fanout-splitting is not allowed, there is no benefit of coding since just routing would suffice. This example shows that the network coding rate region is greater than that of fanout-splitting.

\begin{figure}[h!]
\begin{center}
\includegraphics[width=0.17\textwidth]{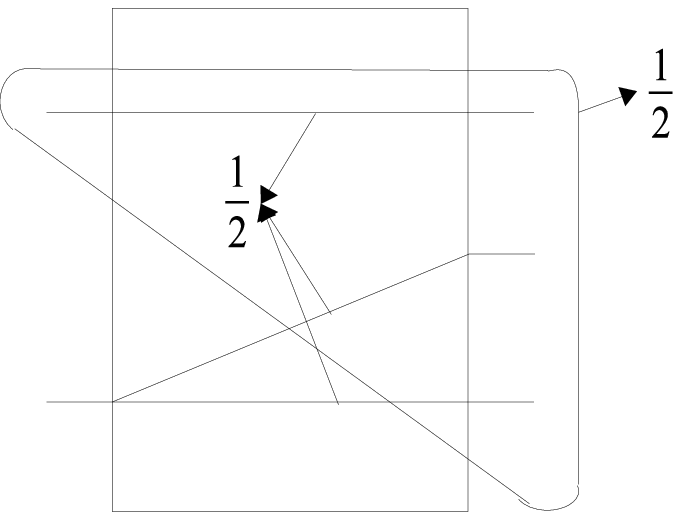}\end{center} \caption{A
traffic pattern which cannot be achieved by network coding}
\label{fig:notachievable}
\end{figure}

However, not all admissible rates are achievable even with network coding. For instance, Figure \ref{fig:notachievable} shows a traffic pattern which is admissible but not achievable even when network coding is allowed. This is because input 2 is fully loaded and thus, needs to serve one of the two unicasts in every slot. As a result, in any slot, input 1 can serve packets to only two outputs. Input 1, thus, requires two slots to serve one packet from its broadcast. Since the broadcast requires a rate of 1/2, input 1 has to serve the broadcast at every time step, leaving no time for its unicast.

This observation brings into attention the question of how much of the admissible rate region does network coding actually
achieve? In Section \ref{sec:networkcodinginswitch}, we shall discuss in more detail such questions regarding the benefit of network coding in switches.

Another class of linear network coding we could consider is \emph{inter-flow coding} \cite{interflow}. Inter-flow coding can encode packets from the same flow as well as packets from different flows that originate from the same input. It can be shown that inter-flow coding has a strictly larger rate region than that of intra-flow coding. However, inter-flow coding is not considered in this work.

\subsubsection{Speedup}\label{sec:speedup}
Multicast traffic patterns such as the one in Figure \ref{fig:notachievable} cannot be sustained even with coding, although they are admissible. To achieve such rate points, the switch needs to provided with some additional capability such as speedup.
\begin{definition}[\textit{\textbf{Speedup}}]
A switch is said to have a \emph{speedup} of $s$ if the switching fabric can transfer $s$ packets over one slot (as defined in Section \ref{sec:multicastswitchmodel}) from an input to an output. This means the switching fabric can go through $s$ configurations within one slot. In other words, during the time it takes for a packet to arrive at the switch on average, the switch can change its configuration $s$ times.
\end{definition}

\begin{figure}[h!]
\begin{center}
\includegraphics[width=0.50\textwidth]{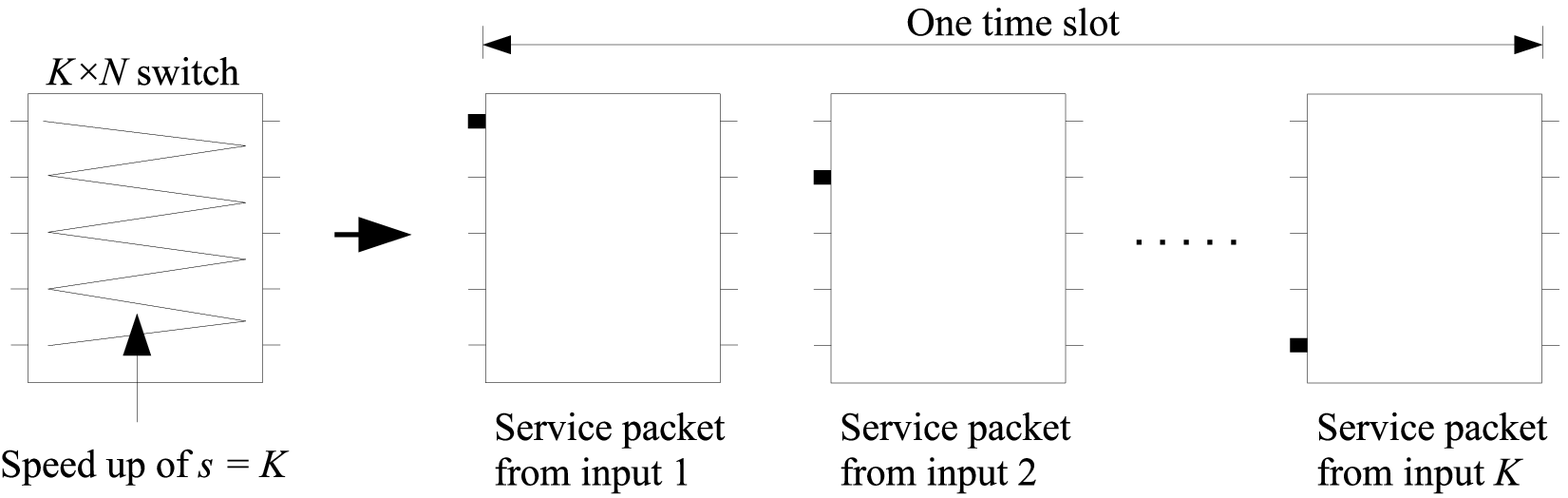}
\end{center}
\caption{Speedup of $s = K$ for an $K \times N$ multicast switch}
\label{fig:speedup}
\end{figure}

It is important to note that with enough speedup, a switch can achieve any multicast traffic pattern even without fanout splitting. For example, in a $K \times N$ switch, if $s \geq K$ then any admissible traffic pattern is achievable. Given any admissible traffic pattern, the switch can divide it up so that each of the $K$ inputs is separately served. Therefore, as shown in Figure \ref{fig:speedup}, the switch will serve whatever traffic input 1 needs to send, then input 2, 3, and so forth. Since the switch has speedup of $s \geq K$, the switch can internally process the $K$ inputs separately and still satisfy all the multicast requirements.

Therefore, a key question is what is the minimum speedup we need to achieve all admissible traffic patterns? From our example in Figure \ref{fig:speedup}, we know that we can upper bound the minimum speedup by $K$ in a $K\times N$ switch even without fanout-splitting or coding; however, can we find a better bound? In addition, as noted in Section \ref{sec:networkcodinginswitch}, we know that network coding increases throughput but not enough to cover the entire admissible rate region. However, we know that with enough speedup any admissible traffic pattern is achievable. Then, our next question is how much speedup does network coding replace? This question will be discussed in more detail in Section \ref{sec:codingvsspeedup}.

\subsection{Graph theory}\label{sec:graphtheory}

In this section, we present some preliminary definitions that will be used throughout this paper. For more detailed and thorough survey on graph theory and combinatorics, see \cite{combinatorics}.

Let $G = (V, E)$ be an undirected graph with vertex set $V$ and edge set $E$. A graph $G_1 = (V_1, E_1)$ is a subgraph of $G$ if $V_1 \subseteq V$ and $E_1 \subseteq E$. A graph $G_2 = (V_2, E_2)$ is an \emph{induced subgraph} of $G$ if $V_2 \subseteq V$ and for all $v_1\in V_2$ and $v_2\in V_2$, we have $(v_1 ,  v_2) \in E_2$ if and only if $(v_1, v_2) \in E$. In addition, $G_2$ is often denoted as $G(V_2)$ and is said to be induced by $V_2$. The \emph{complement} of graph $G$ denoted $\overline{G}$, is a graph on the same vertex set $V$ such that two vertices of $\overline{G}$ are adjacent if and only if they are not adjacent in $G$.

\begin{definition}[\textit{\textbf{Chromatic Number}}]
The \emph{chromatic number} of a graph $G$ is the smallest number of colors  $\chi(G)$ needed to color the vertices of $G$ so that no two adjacent vertices share the same color.
\end{definition}
\begin{definition}[\textit{\textbf{Complete Graph}}]
$G$ is a \emph{complete graph} if for every pair of vertices in $V$ there exists an edge connecting the two.
\end{definition}
\begin{definition}[\textit{\textbf{Multipartite Graph}}]
$G$ is a \emph{multipartite graph} if $V$ can be partitioned into non-empty subsets, called partitions, such that no two vertices in the same partition have an edge connecting them.
\end{definition}
\begin{definition}[\textit{\textbf{Complete Multipartite Graph}}]
$G$ is a \emph{complete multipartite graph} if $G$ is a multipartite graph such that any two vertices that are not in the same partition have an edge connecting them.
\end{definition}
\begin{definition}[\textit{\textbf{Clique}}]
In a graph $G=(V,E)$, a set of vertices $V_1\subseteq V$ is said to form a \emph{clique} if these vertices induce a complete graph.
\end{definition}
\begin{definition}[\textit{\textbf{Clique Number}}]
The \emph{clique number} $\omega(G)$ of a graph $G$ is the number of vertices of the largest clique in $G$.
\end{definition}
\begin{definition}[\textit{\textbf{Stable Set}}]
In a graph $G=(V,E)$, a set of vertices $V_1\subseteq V$ is said to form a \emph{stable set} if for every pair of vertices in $V_1$, there is no edge connecting the two.
\end{definition}
\begin{definition}[\textit{\textbf{Fractional Weighted Coloring Problem}}] Given a graph $G$ and a weight $w_v\in \mathbb{R}^+$ for each vertex, $\text{minimize \ \ \ }\sum_{i=1}^k\lambda_i \hspace{0.3in} \mbox{($\lambda_i \in \mathbb{R}^+, \hspace{0.1in} \forall i$)} $ such that there exist stable sets $\{S_i\}$ of $G$ with $\sum_{i=1}^k\lambda_i\mathbf{\chi}^{S_i}=\mathbf{w}$, where $\mathbf{w}$ is the given weight vector, and $\mathbf{\chi}^S$ denotes the incidence vector of the stable set $S$. The optimum value of the minimization problem is called the \emph{fractional weighted chromatic number}.
\end{definition}
\begin{definition}[\textit{\textbf{Hole}}]
$G$ is a \emph{hole} if it is a chordless cycle; $G$ is called an \emph{odd hole} if it is a hole of odd length at least 5.
\end{definition}
\begin{definition}[\textit{\textbf{Anti-hole}}]
$G$ is an \emph{anti-hole} if its complement is a hole; $G$ is an \emph{odd anti-hole} if its complement is an odd hole.
\end{definition}
\begin{definition}[\textit{\textbf{Perfect Graph}}]
$G$ is said to be \emph{perfect} if for every induced subgraph of $G$, the size of the largest clique equals the chromatic number.
\end{definition}

\subsubsection{Stable set polytope}\label{sec:ssp}
The \emph{stable set polytope} $STAB(G)$ of a graph $G = (V,E)$ is the convex hull of the incidence vectors\footnote{The incidence vector of a set of vertices $V_1\subseteq V$ is a $\{0, 1\}$-vector $\mathbf{x}$ whose entries are labeled with the vertices of $G$. If $x_i = 1$, then vertex $i$ is in $V_1$; otherwise, $i\notin V_1$.} $\mathbf{x}$ of the stable sets of the graph $G$. For a general graph $G$, it is $NP$-hard to compute the stable set polytope $STAB(G)$ and a complete characterization of $STAB(G)$ in terms of linear inequalities is unknown.

However, several families of necessary conditions are known. One example is the \emph{clique inequalities}:
\begin{equation}
\sum_{i\in Q} x_i \leq 1
\end{equation}
for all cliques $Q$ in $G$. Clique inequalities of a graph say that the total weight on the vertices of maximal cliques must not exceed 1. Note that an incidence vector of a stable set must satisfy all the clique inequalities since a stable set can only have at most one vertex from each clique in a graph. Thus, this shows that the clique inequalities are necessary conditions for the stable set polytope. The polytope described by these clique inequalities along with non-negativity constraints
\begin{equation}
x_i \geq 0
\end{equation}
for all nodes $i$ of $G$ is called the \emph{fractional stable set polytope} $QSTAB(G)$. The fractional stable set polytope is often used as a canonical relaxation of $STAB(G)$. Note that, for most graphs, $STAB(G) \subsetneq QSTAB(G)$, since the clique inequalities are necessary but not sufficient conditions for stable set polytope. The two polytopes coincide precisely when $G$ is \emph{perfect}.

Another family of necessary conditions is the \emph{odd hole constraints} \cite{chvatal}:
\begin{equation}
\sum_{i \in H} x_i \leq \left\lfloor \frac{|H|}{2} \right\rfloor
\end{equation}
where $H$ is a set of vertices that induce an odd hole in graph $G$, and $|H|$ denotes the cardinality of $H$. It is easily seen that the incidence vector of a stable set must satisfy the odd hole constraints since a stable set can only have at most one vertex from two adjacent vertices, and therefore, it can include only every other vertex in a cycle.

\subsubsection{Perfect graph}\label{sec:perfect_graph}

From the definitions in Section \ref{sec:graphtheory}, it is not hard to see that in any graph, the clique number is a lower bound on the chromatic number, since all vertices in a clique must be assigned a distinct color in any proper coloring. Perfect graphs are those for which this lower bound is tight for all its induced subgraphs.

One of the important features of perfect graph is that many $NP$-hard graph problems become easy to solve on perfect graphs. For example, the graph coloring problem, maximum clique problem, maximum stable set problems as well as the stable set polytope problems are all known to be solvable in polynomial time for perfect graphs \cite{combinatorics}. In addition, perfect graphs lend us a complete characterization of $STAB(G)$ in terms of linear inequalities: $STAB(G) = QSTAB(G)$ if and only if $G$ is perfect; thus $STAB(G)$ is defined by the clique inequalities and the non-negativity constraints if and only if $G$ is perfect.

We now state three well-known theorems about perfect graphs, which can be found on page 1107 - 1111 of \cite{combinatorics}.

\begin{theorem}\label{thm:weak}\emph{(Weak Perfect Graph Theorem) A graph $G$ is perfect if and only if its complement is perfect.}
\end{theorem}

\begin{theorem}\label{thm:strong}\emph{(Strong Perfect Graph Theorem) A graph $G$ is perfect if and only if it contains no odd hole and no odd anti-hole.}
\end{theorem}

\begin{lemma}\label{thm:replication}\emph{(Replication Lemma) Let $G = (V,E)$ be a perfect graph and $v\in V$. Create a new vertex $v'$ and join it to $v$ and to all the neighbors of $v$. Then, the resulting graph $G'$ is perfect.}
\end{lemma}

Some of the well known perfect graphs that we shall be using in this paper are: complete graphs, bipartite graphs, split graphs (graphs whose vertices can be partitioned into two disjoint sets, which induce a stable set and a clique respectively), and disjoint union of perfect graphs.

It is not hard to imagine that there can be different degree of ``perfection'' in a graph. We can consider two graphs $G$ and $H$ where both are not perfect but $STAB(G)$ and $QSTAB(G)$ are of approximately equal size while $STAB(H)$ is much smaller than $QSTAB(H)$. In such a case, we would consider $G$ to be ``more perfect'' than $H$. This observation gives rise to the need of a metric which measures how perfect a graph is. The \emph{imperfection ratio} \cite{gerke} was introduced precisely for this purpose.

\subsubsection{Imperfection ratio}\label{sec:impratio}

In \cite{gerke}, the imperfection ratio $\imp{G}$ of graph $G$ is defined as
\begin{equation}
\imp{G} = \min\{t : QSTAB(G) \subseteq t\ STAB(G)\}.
\end{equation}
In essence, the imperfection ratio measures how much bigger the fractional stable set polytope $QSTAB(G)$ is relative to the stable set polytope $STAB(G)$. Note that for a perfect graph $G$, $\imp{G} = 1$. Therefore, $\imp{G} \geq 1$ for any graph $G$.

A useful bound on the imperfection ratio is presented in \cite{gerke2} and as Corollary 2.3.5 in \cite{gerkethesis}, which we reproduce below.

\begin{proposition}\label{thm:imp} \emph{(Gerke and McDiarmid)
For a graph $G$, if each vertex in $G$ can be covered $q$ times by a family of $p$ induced perfect subgraphs, then $\imp{G} \leq \frac{p}{q}$.}
\end{proposition}

We shall later revisit this notion of imperfection of a graph when we study the rate regions of multicast switches in Section \ref{sec:codingvsspeedup} and relate this notion to speedup in switches.

\section{Conflict graphs and network coding}\label{sec:networkcodinginswitch}
In a general network, a link may be configured to one of several possible states, for instance, by an algorithm that computes the schedule or the network code. It is likely that the assignment of states to links are dependent on each other. In Section \ref{sec:cg}, we present a graph-theoretic model to capture this dependence in a general network. In Section \ref{sec:ecg}, we apply this approach to the case of multicast switches with network coding to define the notion of the \emph{enhanced conflict graph}.
We shall use this model in Sections \ref{sec:rateregion} and \ref{sec:codingvsspeedup} to obtain our main result.

\subsection{Conflict graph}\label{sec:cg}
Let $\mathcal{N} = (V, E)$ be a directed acyclic graph which represents a network. The conflict graph $\mathcal{N}' = (V', E')$ is an undirected graph corresponding to the network $\mathcal{N}$, and is constructed as follows:

\begin{itemize}
\item For every link $l \in E$, create a set of vertices $v_{(l,s)}$ in $V'$ so that there is a one-to-one correspondence between all the possible states $s$ of link $l$ and the vertices $v_{(l,s)}$.
\item Connect two vertices $v_{(l, s)}$ and $v_{(l', s')}$ if assigning both state $s$ to link $l$ and state $s'$ to link $l'$ simultaneously is impossible. This implies that there is an edge between all pairs of $v_{(l, s)}$ and $v_{(l, t)}$ where $s \ne  t$ since a link cannot be assigned two different states simultaneously. In more general scenarios, we may need to model conflicts using hyperedges to capture cases where a combination of states may be incompatible while any subset of them could coexist. For instance, given a set of inputs, a node can only output a function of those inputs. Thus, if the output link state is not compatible with the combination of input link states,  we connect the vertices corresponding to those states with a hyperedge.
\end{itemize}

Once we have constructed our conflict graph, a stable set represents a collection of states for links such that there is no conflict, \ie, it is possible to assign the set of states to the links in the network. Thus, a valid configuration in the network corresponds to a stable set, and any achievable rate can be achieved by time-sharing between the stable sets. This means that we can represent the achievable rate region by a convex hull of the stable sets, \ie, the stable set polytope of the conflict graph.

Although this conflict graph formulation is easy to conceptualize, it has been noted in  \cite{conflictgraphs} that the size of a conflict graph grows exponentially with the number of possible states for each link. Furthermore, the problem of computing the stable set polytope of a graph is known to be $NP$-hard as discussed in Section \ref{sec:ssp}. Thus, we do not expect to find an efficient algorithm that computes the schedule, given a set of rates in polynomial time with respect to the size of the network. This motivates us to look into combinatorial and graph-theoretic tools to help us understand the structure of the rate region and exploit this structure to design efficient scheduling algorithms.

\subsection{Enhanced conflict graph}\label{sec:ecg}

The \emph{enhanced conflict graph} is a special kind of conflict graph introduced by \cite{conflictgraphs}, which is used to characterize the rate region of multicast switches using network coding. The enhanced conflict graph $G=(V,E)$ for a traffic pattern is an undirected graph defined as follows:

\begin{itemize}
\item For every \emph{subflow}, create a vertex.
\item Vertices representing subflow $(i, J, j)$ and subflow $(i', J', j')$ are connected if and only if
\begin{itemize}
\item $j=j'$, or
\item $i = i'$ and $J \ne J'$.
\end{itemize}
In other words, vertices are adjacent if and only if they have the same output, or if they are from the same input and they belong to different flows.
\end{itemize}

The enhanced conflict graph is constructed such that the maximal cliques reflect the admissibility condition, which we shall formally state and prove in Section \ref{sec:admissiblerateregion}. To briefly discuss the intuition, the constraint that no input should send more than one unique packet at a time is represented by the edges connecting nodes corresponding to subflows $(i, J, j)$ and $(i, J', j')$ where $J\ne J'$. It is important to note that nodes representing subflows from the same flow, for example $(i, J, j)$ and $(i, J, j')$ where $j \ne j'$, are not adjacent. This is because two subflows from the same flow  can be served simultaneously, since input $i$ can send a single packet that is simultaneously useful to multiple outputs by coding packets together. This will be discussed in more detail in the derivation of the achievable region. The second constraint that no output should receive more than one packet at a time is accounted for by the edges connecting vertices of subflows $(i, J, j)$ and $(i', J', j')$ where $j = j'$.

In addition to encoding the admissibility condition with cliques, the enhanced conflict graph also encodes information about achievable rate regions. A stable set in an enhanced conflict graph represents a set of subflows that can be served simultaneously in a valid switch configuration. For instance, any subset of the subflows that belong to the same multicast flow form a stable set, and they can be served simultaneously.

\begin{example}
An example of an enhanced conflict graph of the traffic pattern shown in Figure \ref{fig:codinggood} is given in Figure \ref{fig:ecgfsimp}.

\vspace*{0.2cm}
\begin{figure}[h!]
\begin{center}
\includegraphics[width=0.30\textwidth]{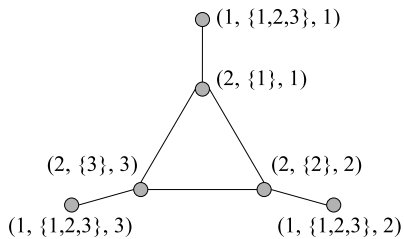}
\end{center}
\vspace{-.2cm} \caption{Enhanced conflict graph of traffic pattern shown in Figure \ref{fig:codinggood}} \label{fig:ecgfsimp}
\end{figure}
\end{example}

This graph-theoretic formulation helps us transform any given traffic pattern in a multicast switch into an enhanced conflict graph, and  the properties of this graph can be used to derive insight on the rate regions of the switch as shown in Section \ref{sec:rateregion}. A similar graph-theoretic formulation was also used by Caramanis \etal \cite{caramanis} in the context of unicast traffic in Banyan  networks.

It is important to note the difference between the enhanced conflict graph and the conflict graphs introduced in Section \ref{sec:cg} or in \cite{caramanis}. In a conflict graph, the vertices represent configurations of a network, and therefore, conflict graphs are not very well equipped to represent fanout-splitting. Reference \cite{caramanis} only considers unicast traffic; therefore, their formulation naturally does not incorporate fanout-splitting. However, the enhanced conflict graphs, by representing subflows with separate vertices, naturally incorporate fanout-splitting capability of a multicast switch.

The enhanced confict graph formulation defined above does not work for the case of fanout splitting without network coding. This is because if coding is not allowed, then it may not always be possible to serve subflows from the same flow, even though the switch allows the input to be connected to multiple outputs. This might happen, for instance, when each output wants a different packet. Without coding, it is not possible to satisfy multiple outputs with a single packet, even if the input is connected to all the outputs.

For the case of fanout-splitting without coding, Marsan \etal \cite{marsan} gave a characterization of the rate region as the convex hull of certain modified departure vectors. However, this formulation does not have a neat graph-theoretic interpretation in general. On the other hand, allowing network coding not only increases throughput, but as we will show in the Section \ref{sec:rateregion}, it also leads to a simpler description of the rate region and enables the use of graph-theoretic tools.

\subsection{Properties of enhanced conflict graph of a multicast switch}\label{sec:propertiesofecg}

In this section, we describe some interesting properties/structure of the enhanced conflict graph of a multicast switch. We show that the class of graphs that are the enhanced conflict graph of some multicast traffic pattern does not cover the class of all possible graphs -- \ie, there is some structure to the enhanced conflict graph. This is of interest from an algorithmic perspective. Since the enhanced conflict graph can be used to characterize the rate region of multicast switches (see Section \ref{sec:rateregion}), it is useful to understand the structure of the enhanced conflict graph in order to determine whether it is possible to develop efficient algorithms to compute the rate region and schedules for multicast switches.

As mentioned in Section \ref{sec:ecg}, in an enhanced conflict graph of a multicast switch, conflicts between a pair of subflows exist due to one or both of the following two reasons:
\begin{itemize}
\item The two subflows go to the same output.
\item The two subflows originate at the same input, and belong to different flows.
\end{itemize}
This constrains the structure of the enhanced conflict graph for multicast switches. First, we make the following observation about subflows arising at the same input.

\begin{figure}[btp]
\centering
\epsfig{file=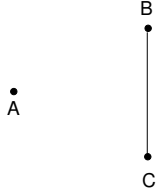, scale=.4}
\caption{A forbidden induced subgraph: co-$P_3$}\label{fig:cop3}
\end{figure}

\begin{lemma}\label{thm:coP3}
\it In the subgraph induced by subflows from the same input, co-$P_3$ is a forbidden subgraph, where co-$P_3$ is given in Figure \ref{fig:cop3}.
\end{lemma}
\IEEEproof
First, we argue that subflows from the same input induce a complete multipartite graph. Consider subflows from each flow at the input as a different partition. By the way edges are defined, subflows from the same flow do not conflict, but any two subflows of two different flows do conflict, resulting in a complete multipartite graph.

A complete multipartite graph cannot contain co-$P_3$ as an induced subgraph. This is because, any two vertices that are not adjacent in a complete multipartite graph must be from the same partition. Referring to Figure \ref{fig:cop3}, vertices $A$ and $B$ must be from the same partition. Similarly, vertices $A$ and $C$ must also be from the same partition. Therefore, $B$ and $C$ must be from the same partition, which means they should not be connected to each other -- a contradiction.
\endproof

\begin{lemma}\label{thm:kneighbors}
\it For any $k>1$, if the enhanced conflict graph has a set $S$ of $k$ vertices such that no two of them are adjacent to each other but they all have a common neighbor $v$, then at least $(k-1)$ of the vertices in $S$ must represent subflows from the same input as $v$.
\end{lemma}
\IEEEproof
Every vertex in $S$ is a neighbor of $v$. So, it either has the same output as $v$, or is from the same input as $v$ but from a different flow. Suppose the given statement is false. Then, at least two vertices in $S$ must represent subflows to the same output as $v$. However, this would imply that these two vertices must be connected to each other, as they conflict at the output, which leads to a contradiction.
\endproof

\subsubsection{Forbidden graphs in the enhanced conflict graph}
\begin{figure}[btp]
\centering
\epsfig{file=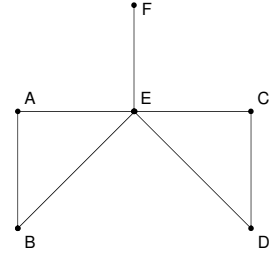, scale=.35}
\caption{The webbed claw}\label{fig:webbedclaw}
\end{figure}

We now present a collection of graphs that can never occur as a subgraph in the enhanced conflict graph of any traffic pattern in a multicast switch.
\begin{theorem}\label{thm:webbedclaw}
\it The webbed claw (shown in Figure \ref{fig:webbedclaw}) cannot occur as an induced subgraph in any enhanced conflict graph.
\end{theorem}
\IEEEproof
Suppose there is a traffic pattern in whose enhanced conflict graph, the webbed claw appears as an induced subgraph. Let $i$ be the input of subflow represented by the vertex $E$ in Figure \ref{fig:webbedclaw}. Then, vertices $B$, $F$, and $D$ are 3 neighbors of $E$ that are not adjacent to each other. By Lemma \ref{thm:kneighbors}, at least two of them must have input $i$. We consider the following cases:

\begin{enumerate}
\item $B$ and $D$ are from input $i$.

Now, $A$ and $C$ are two non-adjacent neighbors of $E$. Again using Lemma \ref{thm:kneighbors}, one of them must represent a subflow from input $i$. Without loss of generality, let this be $A$. Now, $A$, $B$ and $D$ are from the same input and they induce a co-$P_3$. This contradicts Lemma \ref{thm:coP3}.

\item $D$ is not from input $i$.

Then, $B$ and $F$ must both be from input $i$. $A$ and $D$ are two non-adjacent neighbors of $E$. So, by Lemma \ref{thm:kneighbors}, at least one of them is from input $i$. Since $D$ is not from input $i$, $A$ must be from input $i$. Now, $A$, $B$ and $F$ are from the same input and they induce a co-$P_3$. This contradicts Lemma \ref{thm:coP3}.

\item $B$ is not from input $i$.

By symmetry, this is essentially the same as Case 2.
\end{enumerate}

Thus, we get a contradiction in all cases. This completes the proof.
\endproof

\begin{figure}[btp]
\centering
\epsfig{file=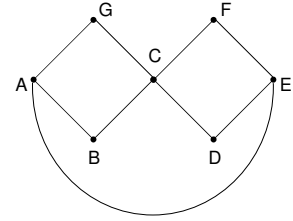, scale=.38}
\caption{The connected double diamond}\label{fig:connecteddoublediamond}
\end{figure}

\begin{theorem}\label{thm:doublediamond}
\it The connected double diamond (shown in Figure \ref{fig:connecteddoublediamond}) cannot occur as an induced subgraph in any enhanced conflict graph.
\end{theorem}
\IEEEproof
Let $i$ be the input of subflow $C$. Now, $B, D, F$ and $G$ are neighbors of $C$, no two of which are connected to each other. Hence, using Lemma \ref{thm:kneighbors}, at least three of them have the same input $i$ as $C$. Without loss of generality, let $B$, $D$ and $F$ be from input $i$.

$D$ and $F$ are non-adjacent neighbors of $E$. Hence, again by Lemma \ref{thm:kneighbors}, at least one of them has the same neighbor as $E$. But, both $D$ and $F$ have input $i$. This means $E$ must also have input $i$.

Now, $B$, $E$ and $F$ are subflows from the same input $i$ and they induce a co-$P_3$. This contradicts Lemma \ref{thm:coP3}. Therefore, the connected double diamond cannot be an induced subgraph of any enhanced conflict graph.
\endproof

\begin{figure}
\centering
\epsfig{file=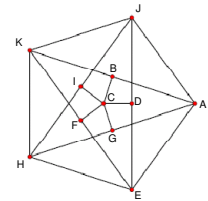, scale=.4}
\caption{The Gr\"otzch graph}\label{fig:grotzch}
\end{figure}

It can be seen that the Gr\"otzch graph (shown in Figure \ref{fig:grotzch}) contains the connected double diamond as an induced subgraph -- vertices $A$ to $G$ induce a connected double diamond. This implies the following.
 \begin{corollary}\label{thm:grotzch}
\it The Gr\"otzch graph cannot occur as an induced subgraph in any enhanced conflict graph.
\end{corollary}
\IEEEproof
Therefore, this theorem follows from Theorem \ref{thm:doublediamond}.
\endproof

It is interesting to note that the Gr\"otzch graph is the third graph ($G_2$) in a sequence of graphs, called the Mycielski graphs \cite{comparingimperfection}. Mycielski graphs $G_0$, $G_1$, $G_2$, ... form a series of graphs with $\omega(G_i) = 2$ for all $i$, but $\chi(G_i) = 2+i$. In addition, $G_i$ contains $G_{i-1}$ as an induced subgraph.

\begin{corollary}\label{thm:mycielski}
\it Mycielski graphs $G_i$, for all $i \geq 2$, cannot occur as an induced subgraph in any enhanced conflict graph.
\end{corollary}

Mycielski graphs are used to prove that there is no upper bound on the imperfection ratio of a general graph \cite{comparingimperfection}. Mycielski graphs form a sequence with unbounded imperfection ratio, \ie, $\imp{G_i}\rightarrow \infty$ for $i \rightarrow \infty$. Therefore, the fact that the enhanced conflict graph does not contain the Mycielski graphs means that we cannot yet rule out the possibility that the imperfection ratio of the enhanced conflict graph may be bounded as the size of the switch grows.

\section{Rate region of a multicast switch}\label{sec:rateregion}
In the next few subsections, we discuss how the stable set polytope of the enhanced conflict graph is related to the rate region of the switch.

Let $\mathbf{r} \in \mathbb{R_+}^f$ be the \emph{rate vector} of a traffic pattern that has $f$ flows. We call the collection of all achievable and admissible rate vectors as the \emph{achievable rate region} $\mathbf{R} \subseteq \mathbb{R_+}^f$ and \emph{admissible rate region} $\mathbf{A} \subseteq \mathbb{R_+}^f$  respectively. For $\mathbf{r} \in \mathbf{R}$, we can construct a switch schedule, which can be viewed as a time sharing between valid switch configurations (\ie, rate decomposition).

Suppose that the total number of subflows in the traffic pattern $\mathbf{r}$ is $m$. Then, the \emph{enhanced rate vector} $\mathbf{e}(\mathbf{r}) \in \mathbb{R}^m$ corresponding to $\mathbf{r}$ is  defined as:\vspace*{-0.2cm}
\[
\vspace*{-0.2cm} \mathbf{e}_{iJj}(\mathbf{r}) = \mathbf{r}_{iJ},
\text{ for all } j\in J.
\]
Therefore, enhanced rate vector is just an extended version of the rate vector so that each flow is duplicated as many times as the number of its subflows. We use the enhanced rate vector as \emph{weights} for vertices of the enhanced conflict graph. As mentioned in Section \ref{sec:ecg}, subflows that can be served simultaneously are not adjacent. Therefore, in an enhanced conflict graph, a valid switch configuration corresponds to a stable set, and a switch schedule corresponds to a convex combination of stable sets of the enhanced conflict graph $G$. This allows us to draw a connection between the stable set polytope of the enhanced conflict graph and the rate regions of the multicast switch.

\subsection{Admissible rate region of a multicast switch}\label{sec:admissiblerateregion}
In this section, we draw a connection between the fractional stable set polytope $QSTAB(G)$ of the enhanced conflict graph $G$ and the admissible rate region of the multicast switch. For a general graph, a complete characterization of the stable set polytope in terms of linear inequalities is unknown. However, the fractional stable set polytope $QSTAB(G)$, a canonical relaxation of $STAB(G)$, can be described by the clique inequalities along with non-negativity constraints, as mentioned in Section \ref{sec:ssp}.

\begin{theorem}\label{thm:adm=qstab}
\it For any rate $\mathbf{r}\in \mathbf{A}$, the enhanced rate vector $\mathbf{e}(\mathbf{r}) \in QSTAB(G)$, \ie, if a non-negative rate vector $\mathbf{r}$ satisfies the admissibility conditions, then its enhanced rate vector $\mathbf{e}(\mathbf{r})$ satisfies the clique inequalities of the enhanced conflict graph $G$.
\end{theorem}
\IEEEproof
Consider any maximal clique $C$ in $G$. Then, by the construction of the enhanced conflict graph, the vertices in $C$ represent subflows that all start from the same input, or all end at the same output, or both. We prove this statement below in two cases:$|C|=2$ and $|C|>2$.

Consider the case where $|C| = 2$, \ie, $C = \{v_1, v_2\}$, where $v_1$ and $v_2$ are vertices of $G$ corresponding to subflows $(i_1, J_1, j_1)$ and $(i_2, J_2, j_2)$ respectively. By construction of the enhanced conflict graph $G$, one of the following must be true: either $i_1 = i_2$ or $j_1 = j_2$. This proves the statement. Therefore, we only need to consider $|C| > 2$.

Now consider two vertices $u$ and $u' \in C$ such that they represent subflows $(i, J, j)$ and $(i', J', j')$ respectively where $i = i'$ or $j = j'$ but not both. (If such a pair of vertices $u$ and $u'$ does not exist in $C$, then $C$ only includes vertices that start from the same input as well as end at the same output, making the statement trivially true.) Suppose $i = i'$ but $j \ne j'$. For any other vertex $v \in C$, $v$ must conflict with both $u$ and $u'$ since $C$ is a clique. Now, since $u$ and $u'$ have different outputs, $v$ can have an output side conflict with at most one of them. Therefore, $v$ must have an input-side conflict with $u$ and $u'$. This implies that all vertices in $C$ represent subflows starting from the same input $i$. A similar argument holds for the case of $i \ne i'$ but $j = j'$. In this case, all subflows in $C$ will have the same output.

Thus, for any maximal clique $C$ in $G$, the vertices in $C$ represent subflows that all start from the same input, or all end at the same output. Now, if $\mathbf{r}\in \mathbf{A}$, then no input or output is overloaded, \ie, the sum of rates of flows starting from a given input or destined to reach a given output must be less than 1. By the way the enhanced conflict graph was defined, at most one subflow from a given flow can be part of a clique. Therefore, the admissibility condition implies the clique inequalities. The non-negativity conditions of $\mathbf{r}$ carry over to $\mathbf{e}(\mathbf{r})$. Thus, the enhanced rate vector $\mathbf{e}(\mathbf{r})$ is in $QSTAB(G)$.
\endproof

Thus, $QSTAB(G)$ corresponds to the admissible rate region of the multicast switch, \ie, $\mathbf{A}$ is a projection of $QSTAB(G)$.

\subsection{Achievable rate region of a multicast switch}\label{sec:achievablerateregion}
In this section, we will establish the achievable rate region in a multicast switch, in terms of the enhanced conflict graph of the underlying traffic pattern.

We first present a few definitions. Since we only consider linear coding across packets of the same flow (intra-flow coding), the state of knowledge of a switch input or output with respect to a particular flow can be represented as a vector space, and the backlog of knowledge between an input and output can be represented as a \emph{virtual queue}, in the same way as described in \cite{ARQforNC}. We restate these definitions here for completeness.

The vector of coefficients used in the linear combination of packets summarizes the relation between the coded packet and the original stream. For a given flow, a node can compute any linear combination whose coefficient vector is in the linear span of the coefficient vectors of previously received coded packets from that flow. Thus, the state of knowledge of a node with respect to a flow can be defined as follows.

\begin{definition}[\textit{\textbf{Knowledge of a node}}]
The \emph{knowledge of a node} with respect to a particular flow at some point in time is the set of all linear combinations of the original packets of that flow that the node can compute, based on the information it has received up to that point. The coefficient vectors of these linear combinations form a vector space called the \emph{knowledge space} of the node, with respect to that flow.
\end{definition}
\begin{definition}[\textit{\textbf{Innovative Packet}}]
A packet transmitted from an input to an output is said to be \emph{innovative} if it conveys previously unknown information to the output. For linear coding, this means that the coefficient vector of the packet is linearly independent of coefficient vectors of all coded packets of that flow received previously by the output, thereby conveying a new degree of freedom. In other words, the coefficient vector is outside the output's knowledge space for the corresponding flow.
\end{definition}

Associated with every subflow is a virtual queue that represents the backlog of knowledge between the input and the output with respect to the corresponding flow. The formal definition is as follows.
\begin{definition}[\textit{\textbf{Virtual Queue}}]
The size of the virtual queue associated with the subflow $(i,J,j)$ is equal to the difference between the dimension of the knowledge space of input $i$ and that of output $j$ with respect to the flow $(i,J)$.
\end{definition}

From this definition, it follows that an arrival to a subflow's virtual queue occurs when a packet arrives into the corresponding flow's physical queue. This also means that an arrival rate vector $\mathbf{r}$ for the flows translates to a rate vector of $\mathbf{e}(\mathbf{r})$, the enhanced rate vector of $\mathbf{r}$, for the virtual queues. A departure (or service) occurs when an innovative packet is conveyed for that subflow. Thus, the size of the virtual queue represents the number of degrees of freedom that still need to be conveyed to the output, in order to communicate all the packets that have arrived so far.

\subsubsection{The scheduling and coding strategy}
We will consider frame-based schedules. A \emph{frame} refers to a set of $F$ consecutive slots, where $F$ is the frame size. Frame-based schedules are schedules that can be specified by a sequence of $F$ switch configurations such that the switch cycles through these configurations periodically. We also call these offline schedules, since the schedule is decided based on prior knowledge of the arrival rates of the flows, and does not use the instantaneous queue size information to decide the switch configuration. We begin with a theorem that provides service guarantees for a certain set of rate vectors. More specifically, we show that in each frame, every queue receives enough service opportunities to match the arrival rate.

Let $q_{iJ}(n)$ be the size of the physical queue of flow $(i,J)$ at the end of the $n^{th}$ frame. Let $a_{iJ}(n)$ denote the number of arrivals into the queue of flow $(i,J)$ during the $n^{th}$ frame. Without loss of generality, we assume that the rates of all the flows are rational.
\begin{theorem}\label{thm:frame}
\it Consider a traffic pattern with a rate vector $\mathbf{r}$ and an enhanced rate vector $\mathbf{e}$. Suppose fanout splitting and linear network coding are allowed. Then, the following statements are equivalent:
\begin{enumerate}
\item $\mathbf{e}\in STAB(G)$, where $G$ is the enhanced conflict graph of the traffic pattern.
\item There exists a coding scheme and a frame-based schedule with a frame size $F$ such that for every flow $(i,J)$, $r_{iJ}F$ is an integer, and the oldest $r_{iJ}F$ packets that were in the flow's queue at the end of frame $(n-1)$ are served by the end of frame $n$, for all $n\ge 1$. If there were fewer than $r_{iJ}F$ packets, then all of them are served.
\end{enumerate}
(Here, `served' means that these packets are conveyed to all outputs in the fanout of the flow and removed from the flow's queue.)
\end{theorem}

\IEEEproof
{\bf Proof of $1\Rightarrow 2$:} We will present a schedule and a coding scheme that ensure that the queues are served as in the theorem statement. In our scheme, the arrivals during a frame are not processed till the beginning of the next frame. The proof is by induction on the frame number $n$.

{\it Basis step:} The queues are assumed to be empty initially, hence there are no packets at the end of frame 0 and the requirement is trivially satisfied for $n=1$.

{\it Induction hypothesis:} Assume the property holds for frame $k$, for all $1\le k\le n$.

{\it Induction step:} Consider frame $(n+1)$. By hypothesis, $\mathbf{e}\in STAB(G)$. So, we can express $\mathbf{e}$ as a convex combination of the incidence vectors of stable sets of the graph:
\[\mathbf{e}=\sum_{i=1}^m \phi_i \chi^{S_i},\]
where $\chi^{S_i}$ denotes the incidence vector of the stable set $S_i$, $\sum_i \phi_i = 1$ and $\phi_i \geq 0$ for all $i$.

Since all rates are assumed to be rational, we can always pick a frame size $F$ such that $r_{iJ}F$ is an integer for all flows. Assuming the $\phi_i$'s are rational, we can choose $F$ such that $\phi_iF$ is also an integer for all $i$.
Using this $F$ as the frame size, we construct a frame-based schedule by appropriate time-sharing among the different switch configurations represented by the stable sets. Thus, out of $F$ slots in a frame, the switch is configured to stable set $S_i$ for $\phi_iF$ slots, for each $i$. In each slot, the stable set specifies to which outputs each input is to be connected, and which flow is meant to be served over that connection.

This schedule has the property that for each flow $(i,J)$, each output $j$ in its fanout set $J$ is connected to input $i$ in exactly $r_{iJ}F$ slots during one frame. However, owing to fanout splitting, different outputs in $J$ may be connected in different sets of slots, with possible overlap. Of the $F$ slots in the schedule, let $T_{iJ}$ be the total number of slots when input $i$ is connected to at least one of the outputs in $J$, for serving flow $(i,J)$. Thus, $T_{iJ}$ is in general more than $r_{iJ}F$ due to fanout splitting.

We propose a coding scheme that uses a maximum distance separable (MDS) code \cite{lincostello}. The key property of an MDS code that we use here is that an $(n,k)$ MDS code can correct up to $(n-k)$  erasures, each of which may occur anywhere in the codeword. Hence, using any $k$ codeword symbols one can retrieve all the information.

In order to guarantee the service of the queues as in the theorem statement, the algorithm must serve the oldest $\min\left(q_{iJ}(n), r_{iJ}F\right)$ packets from the queue of flow $(i,J)$, for each flow. In our coding scheme, the input uses a $(T_{iJ}, r_{iJ}F)$ MDS code.

The information word has $r_{iJ}F$ symbols (packets) and is chosen as follows. If $q_{iJ}(n)\ge  r_{iJ}F$, then the oldest $r_{iJ}$ packets are chosen as the information word. If $q_{iJ}(n)< r_{iJ}F$, then the oldest $q_{iJ}(n)$ packets are chosen along with $ r_{iJ}F-q_{iJ}(n)$ dummy all-zero packets, to form the information word. The input computes the MDS codeword treating these $r_{iJ}F$ packets as symbols of the information word.  The resulting codeword symbols are sent at each of the $T_{iJ}$ transmission opportunities.
Since each output in the fanout of $(i,J)$ is guaranteed to receive $\mathbf{r}_{iJ}F$ codeword symbols, it can retrieve all the $r_{iJ}F$ packets in the information word. The schedule and the code are computed offline, and are known to all inputs and outputs. This proof assumes a mechanism that helps outputs to identify dummy packets that may have been added while forming the information word.

{\bf Proof of $2\Rightarrow 1$:}
This proof was given in~\cite{ucsdita}, and is summarized here for completeness. Suppose there is a frame-based schedule of switch configurations and an associated code such that the requirement in statement 2 of the theorem is met. Consider an arbitrary flow $(i,J)$. Satisfying the requirement in statement 2 implies that when there are $r_{iJ}F$ or more packets in the queue at the beginning of a frame, the schedule is capable of conveying to every output $j\in J$, $r_{iJ}F$ innovative packets or degrees of freedom from that flow by the end of the frame.

Based on this achieving schedule, form $F$ indicator vectors, one for each slot. Each vector has one entry for every subflow such that, the entry is a 1 if the schedule conveys an innovative packet for that subflow in that slot, and 0 otherwise. These vectors can be viewed as indicators for whether each virtual queue received a service or not in that slot. Adding these indicator vectors over all the $F$ slots must then give $F$ times the enhanced rate vector, since the requirement is satisfied for every subflow. In other words, $\mathbf{e}$ is the average of such indicator vectors over all the $F$ slots in the schedule. But, if a set of subflows receive an innovative packet in the same slot, then they must first of all, be conflict-free in terms of the switch constraints, \ie, each indicator vector has to be the incidence vector of some stable set of the enhanced conflict graph. Therefore, $\mathbf{e}$ can be written as a convex combination of the stable sets. This proves that statement 2 implies statement 1.
\endproof

In the above proof, the schedule ensures that each input gets to talk to each output for enough fraction of time about each flow. To make sure that every transmission opportunity is used to convey a new degree of freedom, we need to use an appropriate code. One way to do this is the MDS code idea described in the proof. However, in general, for an $(n,k)$ MDS code to exist, we need to work over a large field size, comparable to $n$.

Since we view the packets as symbols over a finite field while computing the code, the field size is a parameter of interest. Using an MDS code might require a field size that depends on the length of the schedule, which is not desirable. If the field is too large, then we may need more than one packet to represent a single field element, which makes the implementation more difficult. On the other hand, the field should be large enough to ensure that every transmission conveys an innovative packet to all the recipients whenever possible. We will now show an alternate coding strategy that avoids the large field size requirement of MDS codes, and yet achieves the desired innovation guarantee. This coding scheme is based on earlier results of~\cite{tracey} and \cite{jaggi} on multicasting using network coding. Using this approach, for reasonable assumptions on the switch size and the packet size, the field size required will be such that a field element will indeed fit within one packet.

\begin{proposition}\label{fieldsize}
\it In the proof of Theorem~\ref{thm:frame}, a field size equal to the maximum fanout size is sufficient to ensure that every transmission is innovative to all recipients, except those that have already received the packets that were in the queue at the beginning of the frame.
\end{proposition}
\IEEEproof
We use the same notation as in the proof of Theorem \ref{thm:frame}. Consider a network with three layers of nodes. The first layer has a single node -- the source. The second layer nodes correspond to those time-slots in the frame in which flow $(i,J)$ is being served. Thus, there are $T_{iJ}$ such nodes. In the third layer, there is one node corresponding to each output in the fanout of flow $(i,J)$. The source node is connected to all nodes in the second layer. A ``slot-node'' in the second layer is connected to those ``output-nodes'' of the third layer which are served in the corresponding time-slot. All links have unit capacity. Consider the single source multicast problem with network coding, from the source node to all nodes of the third layer. Since the schedule guarantees that every output receives $r_{iJ}F$ transmissions, this means the min-cut of this network is $r_{iJ}F$. Therefore, using the results of~\cite{tracey} and \cite{jaggi}, $r_{iJ}F$ packets can be transmitted to each output using network coding, and the field size required is equal to the number of destinations, which in our case is the size of the fanout. The network coding solution to this new network naturally leads to the code for the switch.
\endproof

\subsubsection{The rate region}\label{sec:stability}
The schedule used in the above proof suggests the following algorithm -- after every $F$ slots, remove $r_{iJ}F$ packets from each queue $(i,J)$ and serve them over the next $F$ slots using an MDS code. This algorithm, viewed at the time-scale of frames (rather than slots), guarantees deterministic service to each queue with a rate of $r_{iJ}F$ packets per frame. Essentially, it ensures the following evolution for the queue of flow $(i,J)$:
\[q_{iJ}(n+1)=\left(q_{iJ}(n)- r_{iJ}F\right)^+ +a_{iJ}(n)\]
Working at the level of frames, we use the above theorem to establish the rate region for a multicast switch with fanout splitting and network coding, under fairly general assumptions on the arrival process.

We use the same assumptions on the arrival process as in Definition 3.4 of~\cite{neely}:
\begin{itemize}
\item $\lim_{t\rightarrow \infty} \frac1t \sum_{\tau=0}^{t-1}E\{a_{iJ}(\tau)\}=r_{iJ}.$
\item $E[a_{iJ}(t)^2|H(t)]\le A_{max}^2$ for all frames $t$, where $H(t)$ represents the history up to frame $t$.
\item For any $\delta>0$, there exists $T$ such that for any $t_0$,

$E\left[\frac1T \sum_{k=0}^{T-1}a_{iJ}(t_0+k|H(t_0))\right]\le r_{iJ}+\delta	$
\end{itemize}

The type of stability we consider is also the same as in Chapter 3 in \cite{neely}, \ie, strong stability -- a queue is strongly stable if it has a finite time average expected backlog.

\begin{definition}[\textit{\textbf{Rate region}}]
For a given traffic pattern, a rate vector $\mathbf{r}$ is said to be \emph{achievable} if there exists a schedule and a coding scheme that ensure that all the virtual queues are strongly stable. The set of all achievable rate vectors of a given traffic pattern is called the \emph{rate region} for that pattern.
\end{definition}

In Lemma 3.6 of~\cite{neely}, the necessary and sufficient conditions for the strong stability of a single queue under admissible arrival and service processes are given. Applying those results in the present context, we arrive at the following result.

\begin{corollary}\label{thm:rateregion}
\it The rate region $\mathbf{R}$ with linear network coding is given by the set of all rate vectors $\mathbf{r}$ such that, the enhanced rate vector $e\mathbf{(r)}\in STAB(G)$ where $G$ is the enhanced conflict graph.
\end{corollary}

Given the set of rates of the various flows in an achievable traffic pattern, the switch schedule can be obtained using a graph-theoretic approach that is discussed further in Section \ref{sec:offline}. For an arbitrary pattern, this is likely to be a hard problem, as it involves certain coloring problems on the enhanced conflict graph. As mentioned in Section \ref{sec:ssp}, a complete characterization of $STAB(G)$ in terms of linear inequalities is unknown for a general graph $G$.
Note that, for most graphs, $STAB(G) \subsetneq  QSTAB(G)$, since the clique inequalities are necessary but not sufficient conditions for a stable set polytope. Thus, the admissible  region is often a strict superset of the achievable rate region, which implies that it is not possible to achieve 100\% throughput even  with coding -- we need speedup. We shall use this connection between the conflict graph and rate regions to draw insights into what kind  of benefit network coding gives us in terms of speedup in Section \ref{sec:codingvsspeedup}.

\section{Network coding and speedup}\label{sec:codingvsspeedup}
In this section, we study the effect of allowing network coding on the speedup requirement in multicast switches. From Section \ref{sec:speedup}, we know that a switch is said to have a \emph{speedup} $s$ if the switching fabric can transfer packets at a rate $s$ times the incoming and outgoing line rate of the switch. This means the switching fabric can go through $s$ configurations within one slot. Given a traffic pattern, an important quantity of interest is the minimum speedup required to sustain all admissible rates, \ie, to achieve 100\% throughput. We denote this as the $s_{\min}$ for that traffic pattern. From the definition of speedup, it is easy to see that a rate vector $\mathbf{r}$ is achievable with speedup $s$ if and only if it is admissible and $\frac{1}{s}\mathbf{r}$ is within the achievable rate region. Using this fact, $s_{\min}$ is simply the smallest value of $s$ such that $\frac{1}{s}\mathbf{r}$ is within the rate region for all admissible rates $\mathbf{r}$. If we denote the admissible and achievable rate regions as $\mathbf{A}$ and $\mathbf{R}$ respectively, then $s_{\min} = \min\{ s\ |\  \mathbf{A} \subseteq s\ \mathbf{R}\}$.

The section is organized as follows. We first present a special traffic pattern for which the value of $s_{\min}$ is lower bounded by around 1.5 without coding, but is exactly 1 (\ie, no speedup) with coding. Then, we present a graph-theoretic bound on $s_{\min}$ for a general traffic pattern in a $K\times N$ switch. Finally, we present numerical simulation results that quantify the actual benefit of network coding in terms of the rate region and speedup.

\subsection{Network coding reduces speedup required: An example}
\begin{figure}[h!]
\begin{center}
\includegraphics[width=0.250\textwidth]{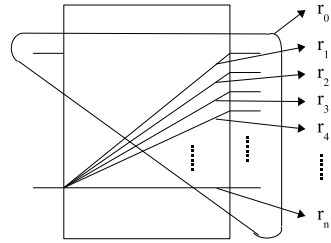}
\end{center}
\caption{A traffic pattern which demonstrates the benefit of coding}
\label{fig:benefit}
\end{figure}

In Figure \ref{fig:codinggood}, we already saw an example of a traffic pattern which can be achieved with network coding but requires a speedup otherwise. In this section, we present a generalization of that example and explicitly quantify the minimum speedup needed to support all admissible traffic rates with and without coding.

The traffic pattern we consider is shown in Figure \ref{fig:benefit}.  It is a $2\times N$ switch with one broadcast flow from input 1 with rate $r_0$ and a unicast from input 2 to every output $i$ with rate $r_i$, for $i\in [N]$. (We use the notation $[m]$ to denote the set of integers from 1 to $m$.)

\begin{figure}[h!]
\begin{center}
\includegraphics[width=0.25\textwidth]{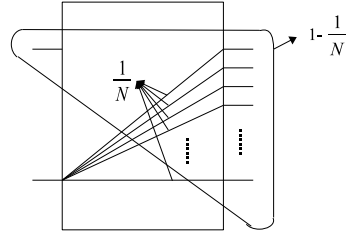}
\end{center}
\caption{A special rate point in the traffic pattern of Figure \ref{fig:benefit}}
\label{fig:benefit1}
\end{figure}

In order to understand the reason for the benefits of coding, we first study a special rate point for this traffic pattern, shown in Figure \ref{fig:benefit1}: set $r_0=\left(1-\frac 1N\right)$ and $r_i=\frac 1N$ for all $i\in [N]$. This means that on average, over a period of $N$ slots, $(N-1)$ packets for the broadcast flow and one packet for each unicast flow must be served. This is clearly an admissible set of rates. It can be seen that Figure \ref{fig:codinggood} corresponds to the special case of $N=3$.

This rate point cannot be achieved with fanout-splitting alone. In every slot, one of the unicasts from input 2 has to be served since input 2 has total inflow of rate 1 and can therefore never be idle. Hence, input 1 needs at least two slots to completely serve each of its broadcast packets. So, it requires at least $2(N-1)$ slots to serve $(N-1)$ packets. This is greater than $N$ for $N>2$. Thus fanout-splitting without coding cannot achieve a rate of $\left(1-\frac 1N\right)$. A speedup is required to achieve this rate point.

On the other hand, this traffic pattern is achievable if network coding is allowed. The schedule is similar to that shown in Figure \ref{fig:codinggood}. During a frame of $N$ slots, input 2 serves the unicasts sequentially starting from output 1 to output $N$ for one slot each, thus achieving the required rate of $\frac{1}{N}$ per unicast. In parallel, input 1 serves the broadcast as follows. In every slot from 1 to $N-1$, it sends a new packet from the broadcast flow to all the outputs except the one occupied by input 2 during that slot. Finally, in the $N^{th}$ slot, it combines all the previous $N-1$ packets using an XOR operation and sends this linear combination to all available outputs. This schedule ensures that output $N$ receives all $N-1$ packets directly. In addition, the remaining outputs $1, 2, \ldots N-1$ also receive enough information to decode all $N-1$ packets. Each of these outputs receives $N-2$ different packets and one XORed packet and can then decode the one remaining packet by applying an XOR operation on all the packets it has received. Thus, $N-1$ packets are delivered over a period of $N$ slots and input 1 successfully completes the broadcast requirement.

Next, we formally quantify the benefit network coding provides compared to fanout-splitting for this specific traffic pattern in terms of the speedup required for achieving all admissible rate points.

To analyze the performance of a network coding switch with the traffic pattern in Figure \ref{fig:benefit}, we present a theorem that identifies a key property of the enhanced conflict graph for this traffic pattern.

\begin{theorem}\label{thm:exampleecg}
\it The enhanced conflict graph for the traffic pattern shown in Figure \ref{fig:benefit} is a perfect graph.
\end{theorem}
\IEEEproof
The enhanced conflict graph consists of a set of $N$ subflows from the broadcast from input 1 at rate $r_0$, and a set of $N$ subflows corresponding to the unicasts from input 2. The unicast subflows form a clique, while the broadcast subflows form a stable set. Thus, the graph is a \emph{split graph}, which is known to be perfect.
\endproof

Now, for a perfect graph $G$, $QSTAB(G)=STAB(G)$. Therefore, comparing Theorem \ref{thm:adm=qstab} and Corollary \ref{thm:rateregion}, we see that the admissible region coincides with the achievable rate region. This leads to the following corollary.

\begin{corollary}\label{thm:examplecor}
\it For the traffic pattern shown in Figure \ref{fig:benefit}, the entire admissible rate region is achievable without any speedup if linear network coding is allowed.
\end{corollary}

Next, we consider the performance of a fanout-splitting switch given the traffic pattern in Figure \ref{fig:benefit}. The rate region of this pattern with fanout-splitting but not coding is given in Theorem \ref{thm:examplerateregion}. ({\it Note: }By rate region, we mean the set of rate vectors for which we can satisfy the same requirement as in statement 2 of Theorem \ref{thm:frame}. The connection to the strong stability of queues can be made in a manner similar to the discussion in Section \ref{sec:stability}.)

\begin{theorem}\label{thm:examplerateregion}
\it The achievable rate region of the pattern shown in Figure \ref{fig:benefit} with fanout-splitting but no coding is given by the following set of inequalities.
\begin{align}
r_i &\geq 0 &\text{for $i = 0, 1, ... N$}\label{eq:nonneg}\\
\sum_{i=1}^N r_i &\leq 1\label{eq:input2}\\
r_0 + r_i &\leq 1 &\text{for $i = 1, 2, ... N$}\label{eq:outputs}\\
2r_0 + \sum_{i=1}^N r_i &\leq 2\label{eq:fanout}
\end{align}
\end{theorem}
The proof is given in the appendix. Note that the conditions (\ref{eq:input2}) and (\ref{eq:outputs}) are the admissibility conditions. The presence of an additional constraint (\ref{eq:fanout}) shows that fanout splitting does not achieve all admissible rates.

We now revisit the special rate point considered earlier: $r_0 = (1-\frac{1}{N})$; $r_i = \frac{1}{N}$ for all $i \in [N]$. Indeed this rate point violates the inequality given in Equation \ref{eq:fanout}, thereby confirming that this point does not lie within the rate region for fanout splitting without coding. The left hand side evaluates to $(3 - \frac{2}{N})$, while the right hand side is only 2. Hence, the smallest scaling factor such that the rate vector lies inside the scaled rate region is $(1.5 - \frac{1}{N})$. This leads to the following corollary.
\begin{corollary}
	\it A speedup of at least $(1.5 - \frac{1}{N})$ is needed to sustain all admissible traffic (\ie, to guarantee 100\% throughput) for the traffic pattern in Figure \ref{fig:benefit} with fanout-splitting but no coding.
\end{corollary}

In other words, we have demonstrated a traffic pattern for which all admissible rates are achievable with no speedup if network coding is allowed, but this needs a speedup of $(1.5 - \frac{1}{N})$ if coding is not allowed. A natural question that follows is -- how much speedup benefit does network coding provide for a general traffic pattern? In particular, does it always achieve all admissible rates?

\subsection{A lower bound on speedup with network coding}
As shown above, network coding can make otherwise unachievable traffic patterns achievable; however, there are admissible traffic patterns that are still unachievable even if we allow network coding. We already presented an example in Figure \ref{fig:notachievable}. We now study this example in greater detail.

\begin{example}
  The traffic pattern in Figure \ref{fig:notachievable} cannot be achieved even when network coding is allowed -- we need other capabilities such as speedup to achieve this traffic pattern. To explain this, we consider the enhanced conflict graph of this traffic pattern as shown in Figure \ref{fig:5/4}. Here, $u_{ij}$ represents the unicast flow vertex from input $i$ to output $j$, and the $b_{ij}$ represents the broadcast subflow vertex from input $i$ to output $j$. The enhanced conflict graph contains an odd hole; hence by Theorem \ref{thm:strong}, it is not perfect. Thus, from Section \ref{sec:networkcodinginswitch}, we know that the achievable rate region is smaller than admissible rate region; the switch needs speedup to achieve this traffic pattern even if we have network coding.

\vspace*{0.5cm}
\begin{figure}[h!]
\begin{center}
\mbox{\includegraphics[width=0.2\textwidth]{config2}}
\mbox{\includegraphics[width=0.23\textwidth]{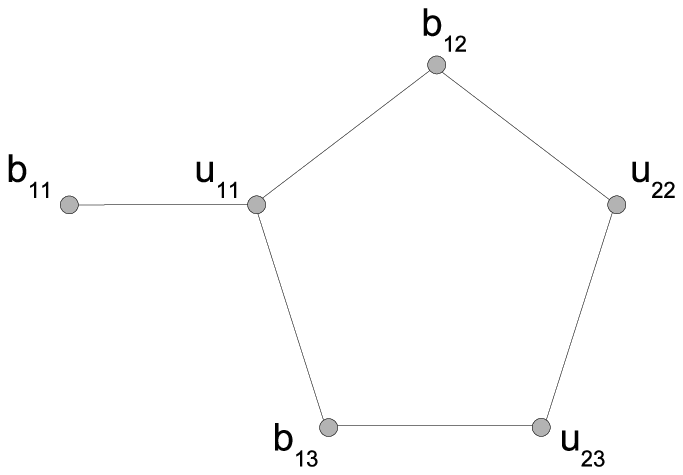}}\end{center} \caption{A
traffic pattern which requires speedup and its enhanced conflict
graph}\label{fig:5/4}
\end{figure}

It turns out that the traffic pattern in Figure \ref{fig:5/4} requires a speedup of 1.25 with network coding. To understand why, we consider the description of the stable set polytope of the enhanced conflict graph. As mentioned in Section \ref{sec:ssp}, there are many necessary conditions for a stable set polytope, such as the odd hole constraints:
\begin{equation}\label{eq:oddholeeq}
\sum_{i\in H} x_i \leq \Bigg\lfloor\frac{|H|}{2}\Bigg\rfloor,
\end{equation}
where $H$ is an odd hole.

\begin{figure*}[btp]
\centering
\includegraphics[width=0.75\textwidth]{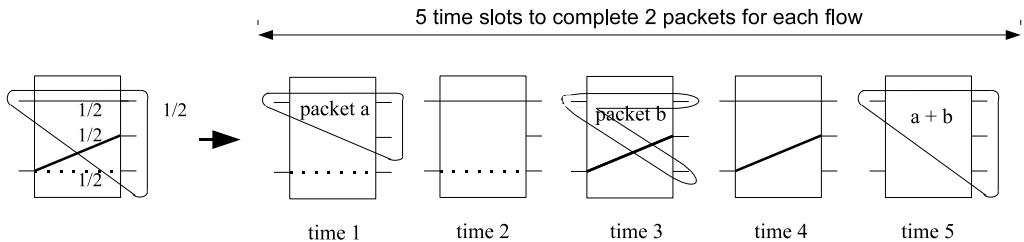} \caption{A traffic pattern that requires speedup in a network coding switch}\label{fig:5/4schedule}
\end{figure*}

We observe that in Figure \ref{fig:5/4} each vertex in the odd hole represents a flow of rate 1/2. Therefore, the total weight on the odd hole is 5/2, which is the total rate the switch needs to serve to satisfy the subflows represented by the vertices in the odd hole.   However, the right-hand side of Equation \ref{eq:oddholeeq} is $\lfloor |H|/2\rfloor = \lfloor 5/2 \rfloor = 2$. Hence, the smallest scaling factor such that the rate vector satisfies the scaled odd hole constraint is $5/4=1.25$. Therefore, a speedup of at least 1.25 is needed to serve this traffic pattern in a network coding switch.

On the other hand, we show that this traffic pattern only requires speedup of at most 1.25 when network coding is allowed. To demonstrate, we present a schedule in Figure \ref{fig:5/4schedule}. Here, the switch serves two packets for each flow. To achieve the required rate of 1/2, this should take 4 slots. However, the switch actually uses 5 configurations. Therefore, the 5 switch configurations have to be mapped to 4 actual slots, which requires a speedup of 1.25. Hence, this shows that the speedup needed to achieve this traffic pattern is exactly 1.25.
\end{example}

In the rest of this section, we seek to quantify the minimum speedup $s_{\min}$ needed to achieve any admissible rate point for an arbitrary traffic pattern in a switch that uses network coding. Note that we already have a lower bound -- the traffic pattern in Figure \ref{fig:5/4} implies that $s_{\min} \geq 1.25$. We will next provide a upper bound on $s_{\min}$.

\subsection{Imperfection ratio bounds speedup}\label{sec:speed_ratio}

This section develops our main result, which relates speedup with imperfection ratio \cite{codingforspeedup}. The key observation here is that if the enhanced conflict graph $G$ is perfect, then by definition $STAB(G) = QSTAB(G)$. In this case, the problem of computing $STAB(G)$ becomes easy, and therefore, computing the achievable rate region of a switch is easy as well. In addition, as noted in Section \ref{sec:impratio}, the less ``imperfect'' a conflict graph is, the closer the stable set polytope is to the fractional stable set polytope. Therefore, imperfection ratio translates to how close the achievable rate region is to the admissible rate region. Thus, understanding and measuring the perfectness of the enhanced conflict graph is a useful way of gaining insight into the benefit of network coding. The relation between the imperfection ratio and the speedup is stated formally below.

\begin{theorem}\label{thm:main}
\it Given a traffic pattern, let $G$ be its enhanced conflict graph and $s_{\min}$ be the minimum speedup required to achieve all admissible rates. Then, \[s_{\min} \leq \imp{G}.\]
\end{theorem}
\IEEEproof
Let $\mathbf{A}$ and $\mathbf{R}$ denote the admissible and achievable rate regions for the given traffic pattern.

$\mathbf{r}\in \mathbf{A}$
\begin{eqnarray*}
&\Rightarrow & \mathbf{e}(\mathbf{r})\in QSTAB(G)\ \ \ \mbox{(Theorem \ref{thm:adm=qstab})}\\
& \Rightarrow & \mathbf{e}(\mathbf{r})\in \imp{G}STAB(G) \ \ \ \mbox{(by definition of $\imp{G}$)}\\
& \Rightarrow & \mathbf{e}\left( \frac 1{\imp{G}} \mathbf{r}\right) \in STAB(G) \ \ \ \mbox{($\mathbf{e}(\cdot)$ is linear)}\\
& \Rightarrow & \frac 1{\imp{G}} \mathbf{r} \in \mathbf{R} \ \ \mbox{(Corollary \ref{thm:rateregion})}\\
\end{eqnarray*}
This implies that $\mathbf{A}\subseteq \imp{G} \mathbf{R}$ and the result follows.
\endproof

Note that the converse of Theorem \ref{thm:main} is not true. This is because enhanced conflict graph $G$ replicates a multicast flow into subflows, and as a result, induces a stable set polytope of dimension greater than the number of actual flows in the traffic. Thus, $\mathbf{A}$ and $\mathbf{R}$ are projections of $QSTAB(G)$ and $STAB(G)$ such that the subflows corresponding to the same  multicast flow have the same weight. As a result, $QSTAB(G) \subseteq \imp{G} STAB(G)$ implies the $\mathbf{A} \subseteq \imp{G} \mathbf{R}$, but $\mathbf{A} \subseteq s_{\min} \mathbf{R}$ may not imply $QSTAB(G) \subseteq s_{\min} STAB(G)$.

\subsection{Bounds on speedup for $K\times N$ switch with unicasts and broadcasts}\label{sec:KxN}

In this section, we apply Theorem \ref{thm:main} to $K\times N$ switches using intra-flow coding with traffic patterns consisting of unicasts and broadcasts only. We show that the minimum speedup needed for 100\%  throughput in this case is bounded by $\min (\frac{2K-1}{K}, \frac{2N}{N+1})$. The rest of this section is organized as follows. First, we give a  description of the enhanced conflict graph for a $K\times N$ switch. In Sections \ref{sec:2k-1/k} and \ref{sec:2n/n+1}, we show the upper and lower bounds on speedup of $\frac{2K-1}{K}$ and $\frac{2N}{N+1}$, respectively.

\subsubsection{Enhanced conflict graph for $K\times N$ switch}\label{sec:ecgKxN}

Consider traffic patterns which consist only of unicasts and a broadcast per each input on a $K\times N$ switch. In such a case, the enhanced conflict graph denoted $G_{K,N} = (V,E)$ has the following structure. (We use the notation $[m]$ to denote the set of integers from 1 to $m$.)

\begin{figure}[h!]
\begin{center}
\includegraphics[width = 0.50\textwidth]{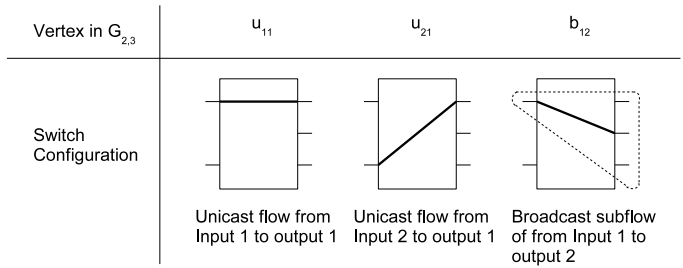}
\end{center}
\vspace{-.2cm} \caption{Switch configuration corresponding to $u_{11}$, $u_{21}$, and $b_{12}$ in $G_{2,3}$}
\label{fig:configuration}
\end{figure}

Each vertex in $G_{K,N}$ represents a subflow in a $K\times N$ switch. The vertex $u_{ij}$ represents the unicast flow from input $i$ to output $j$, and the vertex $b_{ij}$ represents the broadcast subflow from input $i$ to output $j$. As an example, Figure \ref{fig:configuration} shows the switch configuration corresponding to $u_{11}$, $u_{21}$, and $b_{12}$ in a $2\times 3$ switch. Thus, the vertex set is given by \[V = \left(\cup_{i \in [K]} U_i\right) \cup \left(\cup_{i \in [K]} B_i\right)= \left(\cup_{j \in [N]} U^o_j\right) \cup \left(\cup_{j \in [N]} B^o_j\right)\] where
\begin{align*}
 U_i &:= \{ u_{ij}\ |\ j\in [N]\},\ \ \ \ B_i := \{ b_{ij}\ |\ j \in [N]\},\\
 U^o_j &:= \{ u_{ij}\ |\ i \in [K]\},\ \ \ \ B^o_j := \{ b_{ij} |\ i \in  [K]\}.
\end{align*}

Thus, $U_i$ and $U^o_j$ are collections of the unicast flows from input $i$ and to output $j$ respectively. $B_i$ and $B^o_j$ are collections of the broadcast subflows from input $i$ and to output $j$ respectively.

\vspace*{0.5cm}

The intuition behind a conflict  graph is that vertices which represent flows that cannot be served simultaneously are adjacent. Note that if fanout splitting and network coding are allowed, the switch can simultaneously serve two or more subflows of the same broadcast flow and hence such subflows are not adjacent to each other. Hence, the edge set $E = \left( \cup_{i \in [K]} E^u_i\right) \cup \left(\cup_{i \in [K]} E^b_i\right) \cup \left(\cup_{i\in[N]} E^o_i\right)$ where
\begin{align*}
E^u_i &:= \{ (u_{ij}, u_{ik})\ |\ j \ne k; j, k\in [N]\},\\ E^b_i &:= \{ (b_{ij}, u_{ik})\ |\ j, k\in [N]\}, \mbox{ and}\\
E^o_i &:= \{(u_{ji}, u_{ki}), (b_{ji}, b_{ki}), (b_{ji}, u_{ki})\ |\ j \ne k; j,k \in[K]\}\\
\end{align*}

Each edge set represents a different  type of conflict. $E^u_i$ represents conflicts among unicasts at input $i$; $E^b_i$ represents conflict between any broadcast subflow  and any unicast at input $i$; and $E^o_i$ represents conflicts among all flows and subflows at output $i$.

From the input perspective, $G_{K,N}$ consists of $K$ induced complete  subgraphs $G_{K,N}(U_i)$ for unicasts from each input $i$, and $K$ induced stable sets $G_{K,N}(B_i)$ for broadcasts from each input  $i$; from the output perspective, $G_{K,N}$ consists of $N$ induced complete subgraphs $G_{K,N}(U^o_j\cup B^o_j)$ for unicasts and broadcast subflows to output $j$, for each $j\in [N]$.

\begin{example}
For  example, in Figure \ref{fig:2x3}, we show an enhanced conflict graph for a $2\times 3$ switch with unicasts and broadcasts only. There  is an edge between $u_{11}$ and $b_{12}$, since they both represent flows serving input 1. There also exists an edge between $u_{11}$ and  $u_{21}$ since they both serve output 1; however $u_{21}$ and $b_{12}$ are not adjacent since they do not conflict on the input nor the  output side. We can observe that vertices $u_{1j}$ for all $j$, representing unicast flows from input 1, are  adjacent to each other due to input side conflict. This statement holds for $u_{2j}$ for all $j$ as well. Furthermore, we can observe that $b_{1j}$ for all $j$, representing broadcast subflows from input 1, are not adjacent to each other since broadcast subflows from  the same flow can be served simultaneously. Therefore, we can think of $G_{2,N}$ consisting of two induced complete subgraphs  $G_{2,N}(U_1)$ and $G_{2,N}(U_2)$ of size $N$ and two induced stable sets $G_{2,N}(B_1)$ and $G_{2,N}(B_2)$ of size $N$.

\vspace*{0.5cm}
\begin{figure}[h!]
\begin{center}
\includegraphics[width=0.40\textwidth]{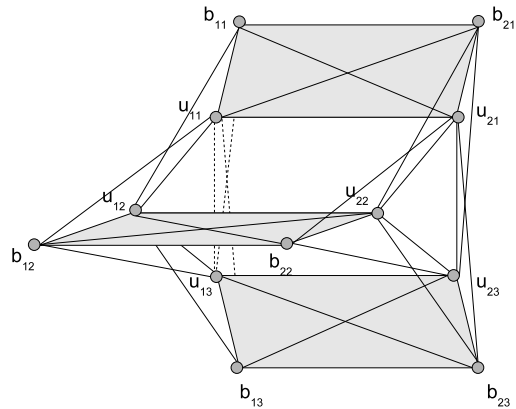}
\end{center}
\vspace{-.2cm} \caption{$G_{2,3}$ for a $2\times 3$ switch with
unicasts and broadcasts only} \label{fig:2x3}
\end{figure}
\end{example}

\begin{figure}[h!]
\begin{center}
\includegraphics[width=0.40\textwidth]{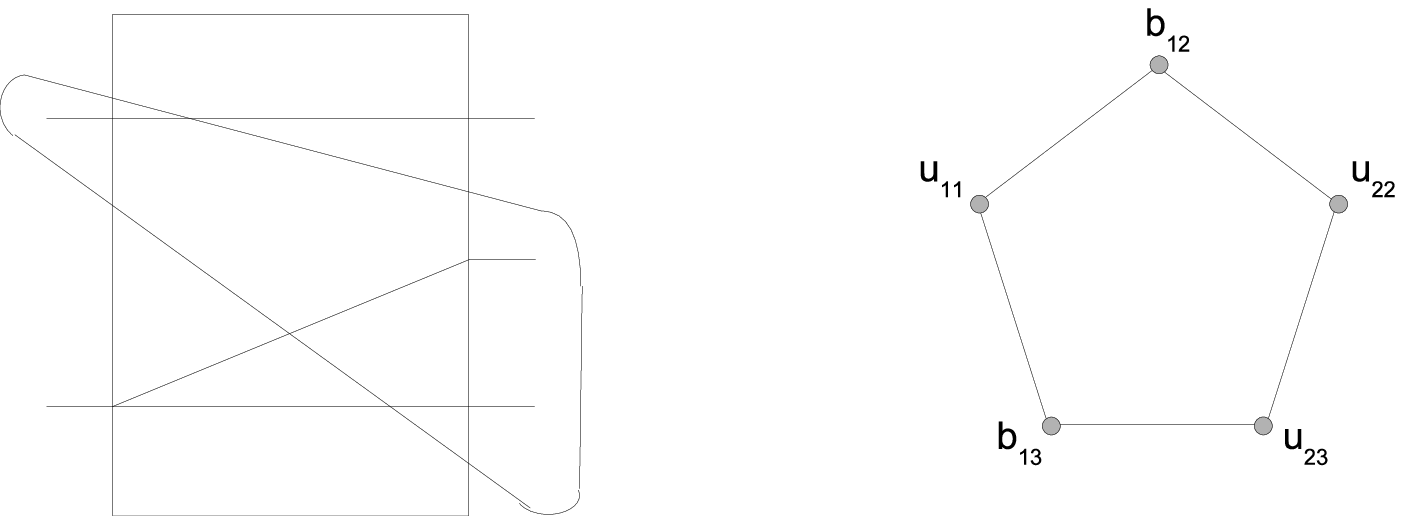}
\end{center}
\caption{A traffic pattern and its enhanced conflict graph}
\label{fig:relax1}
\end{figure}

\begin{figure}[h!]
\begin{center}
\includegraphics[width=0.40\textwidth]{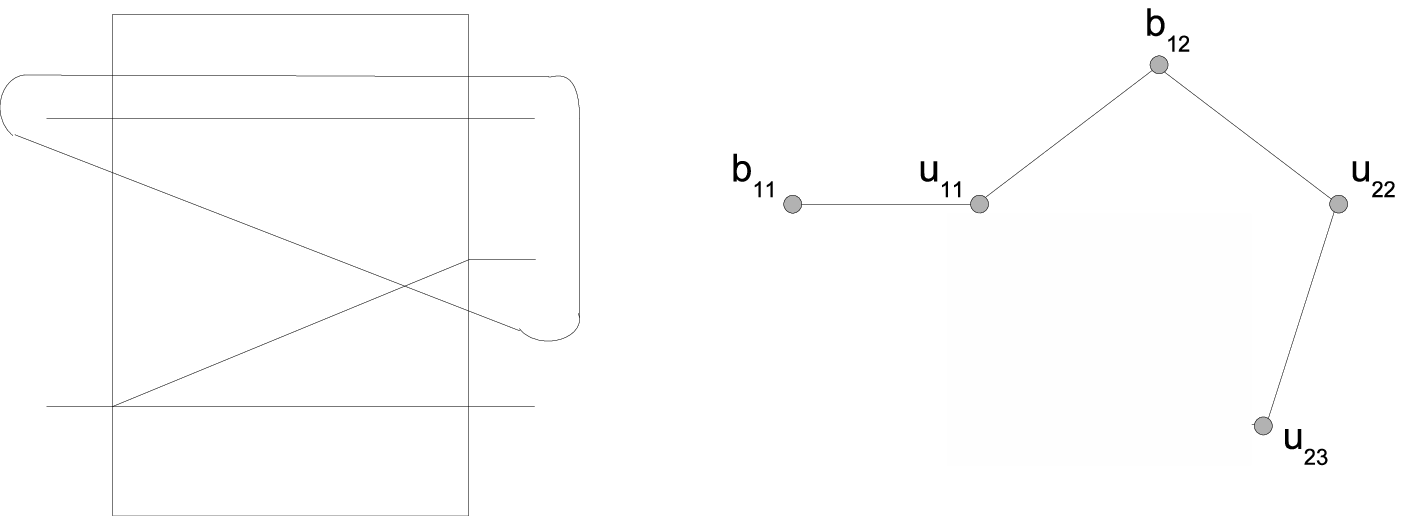}
\end{center}
\caption{A traffic pattern and its enhanced conflict graph}
\label{fig:relax2}
\end{figure}

Here, we note that the conflict graph of a $K\times N$ multicast switch with unicasts and broadcasts can be relaxed to that of unicasts and a single multicast per input. This relaxation just removes vertices that represent broadcast subflows, which are not part of  the multicast flow. Removing vertices from a graph cannot hurt the perfection of a graph.  Therefore, any upper bound on the imperfection ratio of the conflict graph for unicasts and broadcasts bounds holds also for unicasts and a single multicast per input. For example, Figure \ref{fig:relax1} and \ref{fig:relax2} present two traffic patterns which relax the broadcast requirement of the traffic pattern shown in Figure \ref{fig:notachievable}. In Figure \ref{fig:relax1}, input 1 multicasts to only outputs 2 and 3; therefore, the node $b_{11}$ in Figure \ref{fig:notachievable} is removed here. In Figure \ref{fig:relax2}, input 1 multicasts to only outputs 1 and 2; therefore, the node $b_{13}$ in Figure \ref{fig:notachievable} is removed here. The imperfection ratio of the enhanced conflict graph in Figure \ref{fig:relax1} remains the same as that of Figure \ref{fig:notachievable}, since an odd hole of size 5 is present in both. However, in Figure \ref{fig:relax2}, the enhanced conflict graph is perfect, since it is a bipartite graph. This illustrates the fact that removing vertices from a graph cannot make it less perfect.

\subsubsection{Speedup of $\frac{2K-1}{K}$}\label{sec:2k-1/k}
In this section, we give an upper bound on speedup for $K\times N$ switches. We present $2K-1$ induced perfect subgraphs of $G_{K,N}$ that cover all the vertices $K$ times. Then, with Proposition \ref{thm:imp}, we have $\frac{2K-1}{K}$ as an upper bound for speedup.

\begin{lemma}\label{thm:unicasts}
\emph{Let $G^u = G_{K,N}(\cup_{i \in [K]} U_i)$ be an induced subgraph of $G_{K,N}$. Then $G^u$ is perfect.}
\end{lemma}
\IEEEproof
$G^u$ is an enhanced conflict graph for unicast traffic. One may check that $G^u$ is a line graph of a bipartite graph, which is known to be perfect \cite{combinatorics}.
\endproof

Lemma \ref{thm:unicasts} also follows from the result in \cite{100forunicast} which shows that 100\% throughput can be achieved in a  input-queued crossbar switch in the context of unicast traffic.

\begin{lemma}\label{thm:umm}
\emph{Let $G_i = G_{K, N}\left((\cup_{j \in [K]} B_j) \cup U_i\right)$ for some $i \in [K]$ be an induced subgraph of $G_{K,N}$. Then $G_{i}$ is perfect.}
\end{lemma}
\IEEEproof
Assume that $G_i$ is not perfect. So it must have an odd hole or odd anti-hole as an induced subgraph. Suppose it has an odd hole, say $H$. In $G_i$, any broadcast subflow, except the ones from input $i$, has no conflict on the input side. Suppose such a subflow were part of $H$, then both its neighbors in $H$ will be due to output side conflicts. But in that case, the two neighbors will themselves conflict at the output, thereby forming a triangle. Since an odd hole cannot contain a triangle, we conclude that $H$ cannot include any $b_{jk}$ with $j\neq i$.

This means $H$ must be an induced subgraph of $G_{K,N}(B_i \cup  U_i)$. However, $B_i$ induces a stable set, while $U_i$ induces a clique. Therefore, $G_{K,N}(B_i \cup U_i)$ is a split graph, which is known to be perfect. This contradiction shows that $G_i$ cannot contain  an odd hole $H$.

Suppose $G_i$ contains an odd anti-hole. This will happen if and only if $\overline{G_i}$ contains an odd hole, say $H'$. Note that  in $\overline{G_i}$, two vertices are connected if the corresponding subflows do not conflict. Now, $H'$ has to contain at least one unicast, say $u_{ij}$. This is because the broadcasts by themselves induce a perfect subgraph in $\overline{G_i}$, which is a complement of a disjoint union of complete graphs. Now, $u_{ij}$ in $\overline{G_i}$ is adjacent to  any $b_{i'j'}$, where $i\neq i'$ and $j\neq j'$. Let $b_{pq}$ and $b_{p'q'}$ be vertices adjacent to $u_{ij}$ in $H'$. Then, using the  definition of $\overline{G_i}$, we can infer that $i\neq p\neq p'\neq i$ and $q=q'\neq j$. But this means, any vertex that is adjacent  to $b_{pq}$ is also adjacent to $b_{p'q'}$. Hence, $H'$ cannot be an odd hole.

This proves that $G_i$ is perfect.
\endproof

Using Lemmas \ref{thm:unicasts} and \ref{thm:umm}, we derive our first upper bound on speedup in $K\times N$ multicast switches with traffic patterns consisting of unicasts and broadcasts only.

\begin{proposition}\label{thm:2k-1/k}
\emph{$\imp{G_{K,N}} \leq \frac{2K-1}{K}$.}
\end{proposition}
\IEEEproof
Consider the following collection of induced subgraphs: $K-1$ copies of $G^u$ from Lemma \ref{thm:unicasts} and $G_i$ from Lemma \ref{thm:umm} for all $i \in [K]$. We know that these subgraphs are all perfect. In addition, these subgraphs cover each vertex in $v  \in G_{K,N}$ $K$ times. For $v \in U_i$, $G_i$ and each copy of $G^u$ covers $v$ once. For $v \in B_i$, each $G_i$ covers $v$. By Proposition   \ref{thm:imp}, the claim follows.
\endproof

\begin{figure}[h!]
  \centerline{
    \mbox{\includegraphics[width=0.25\textwidth]{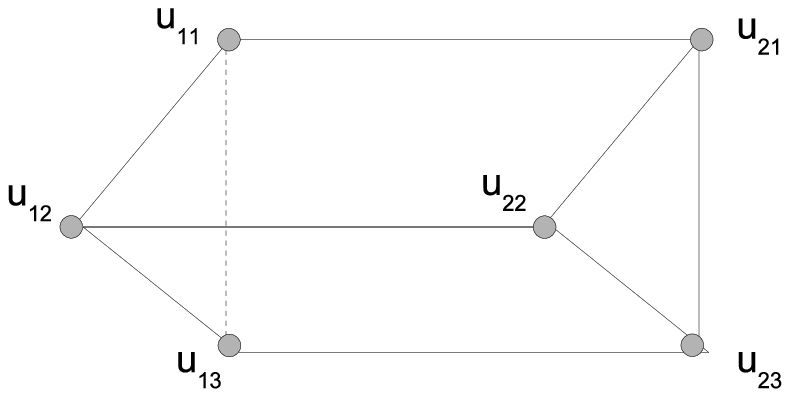}}
    \mbox{\includegraphics[width=0.25\textwidth]{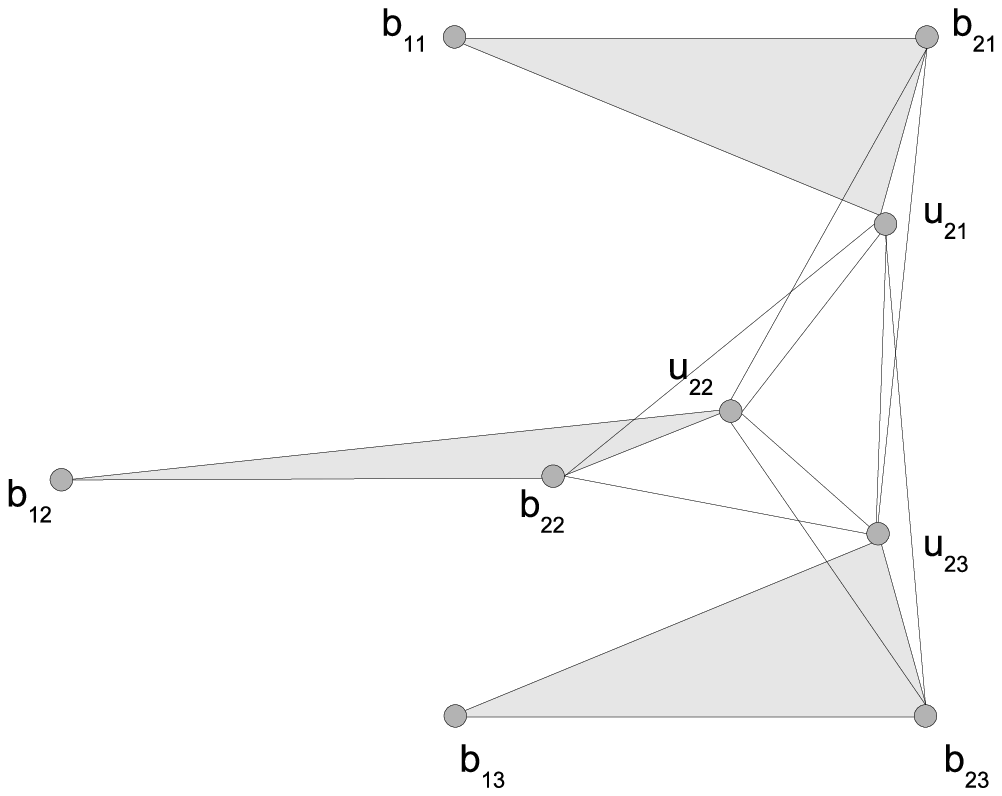}}
  }
  \caption{$G^{u}$ and $G_{2}$ for a $2\times 3$ switch with unicasts and broadcasts only}
  \label{fig:2x3subgraphs}
  \end{figure}

For example, in the $2\times 3$ switch, $G^u$ and $G_2$ for a $2\times 3$ switch is shown in Figure \ref{fig:2x3subgraphs}. In this case, $G^u$, $G_1$ (not shown) and $G_2$ will together cover every vertex in $G_{2,N}$ exactly 3 times. This implies an upper bound of 1.5 on the speedup.

\subsubsection{Speedup of $\frac{2N}{N+1}$}\label{sec:2n/n+1}

The proof idea in this section is similar to that of Section \ref{sec:2k-1/k}. We present $2N$ induced perfect subgraphs of $G_{K,N}$  that cover all the vertices $N+1$ times, and then appeal to Proposition \ref{thm:imp}. However, unlike Section \ref{sec:2k-1/k}, here we  change our focus from the input to output.

\begin{lemma}\label{thm:outunicasts}
\emph{Let $G^o_{1,i} = G_{K,N}(V_i)$ where $V_i = U^o_i \cup \left(\cup_{j \in [N]} B^o_j\right)$ be an induced subgraph of $G_{K,N}$.  Then $G^o_{1,i}$ is perfect.}
\end{lemma}
\IEEEproof
Assume that $G^o_{1,i}$ is not perfect. So it must have an odd hole or odd anti-hole as an induced subgraph. Suppose it has an odd  hole, say $H$. Since $U^o_i \cup B^o_i$ forms a complete graph (known to be perfect), $H$ must contain vertices of $B^o_j$, $j\ne i$.  Suppose $b_{kj} \in B^o_j$ is part of $H$, then $H$ contains at least two vertices of $B^o_j$. This is because, in $G^o_{1,i}$, $b_{kj}$  has only one conflict on the input side; thus, neighbors of $b_{kj}$ are $u_{ki}$ (input conflict) and $B^o_j$ (output conflict).  However, note that $B^o_j$ itself forms a complete graph, therefore $H$ contains at most two vertices of $B^o_j$. Thus, $b_{kj}$ and $b_{k'j}$, $k \ne k'$ are in $H$. Then, $u_{ki}$ and $u_{k'i}$ are in $H$. However, these four vertices form a cycle, thus $G^o_{1,i}$ cannot contain an odd hole $H$.

By the same argument as in the proof for Lemma \ref{thm:umm}, we can show that $G^o_{1,i}$ cannot contain an odd anti-hole. This proves our claim.
\endproof

\begin{lemma}\label{thm:outUUM}
\emph{Let $G^o_{2,i} = G_{K,N}(V_i)$ where $V_i = B^o_i \cup \left(\cup_{j \in [N]} U^o_j \right)$ be an induced subgraph of $G_{K,N}$. Then, $G^o_{2,i}$ is perfect.}
\end{lemma}
\IEEEproof
$G^o_{2,i}$ is an enhanced conflict graph for unicast traffic in addition to all broadcast subflows to output $i$. Consider $b_{1i} \in  B^o_i$ and $u_{1i} \in \cup_{i \in [K]} U_i$. In a $K\times N$ switch, $b_{1i}$ and $u_{1i}$ represent subflows from input 1 to output  $i$, and thus conflict with the same set of subflows, \ie, neighbors of $u_{1i}$ are neighbors of $b_{1i}$. In addition, $b_{1i}$ and  $u_{1i}$ are in conflict. Therefore, by Replication Lemma (Lemma \ref{thm:replication}), we know that $G^o_{2,i}$ is perfect if  $G_{K,N}(V_i \setminus \{b_{1i}\})$ is perfect. We can apply this argument repeatedly for each $b_{ji} \in B^o_i$, and deduce that if  $G_{K,N}(\cup_{j \in [N]} U^o_j)$ perfect then $G^o_{2,i}$ is perfect. Note that from Lemma \ref{thm:unicasts}, we know that the  enhanced conflict graph $G^u = G_{K,N}(\cup_{i \in [K]} U_i) = G_{K,N}(\cup_{j \in [N]} U^o_j)$ for unicast traffic is perfect.  Therefore, $G^o_{2,i}$ is perfect.
\endproof

Now, using Lemmas \ref{thm:outunicasts} and \ref{thm:outUUM}, we can derive an upper bound for speedup in $K\times N$ multicast  switches with traffic patterns consisting of unicasts and broadcasts only.

\begin{proposition}\label{thm:2n/n+1}
\emph{$\imp{G_{K,N}} \leq \frac{2N}{N+1}$.}
\end{proposition}
\IEEEproof
Consider the following collection of induced subgraphs: $G^o_{1,i}$ and $G^o_{2,i}$ for all $i \in [N]$. By Lemmas \ref{thm:outunicasts} and \ref{thm:outUUM}, we know that these subgraphs are all perfect. In addition, these subgraphs cover each vertex  in $v \in G_{K,N}$ $N+1$ times. By Proposition \ref{thm:imp}, the claim follows.
\endproof

\subsection{Numerical study of network coding benefits}\label{sec:sim_improverate}

As noted in Section \ref{sec:background}, we know that network coding increases the throughput of networks in general. We now quantify the benefit of network coding by numerically computing the rate regions. However, as noted in Section \ref{sec:cg}, computing the rate region (which is equivalent to computing the stable set polytope of a conflict graph) is $NP$-hard. As a result, we focus on the rate regions of $2\times 3$ switch with all flows and $4\times 3$ switch with unicasts and broadcasts only.

In a $2\times 3$ switch, there are three unicasts, three two-casts and one broadcast from each of the two inputs. Therefore, the  rate region is a 14-dimensional polytope, which allows numerical computation to be feasible. We computed the stable set polytope of the  enhanced conflict graph corresponding to a $2\times 3$ switch to obtain the different rate regions. The comparison is shown in Table \ref{tab:poly}. In a $4\times 3$ switch with unicasts and broadcasts only, there are three unicasts and one broadcast from each of the four inputs. Therefore, the rate region is a 16-dimensional polytope. We again computed and compared the different rate regions. The results are shown in Table \ref{tab:poly2}.

\begin{table}
\begin{center}\caption{A comparison of the four schemes in a $2\times 3$ switch with all flows}\label{tab:poly}
\begin{tabular}{|l|c|p{1.5cm}|p{1.8cm}|}
\hline
Polytope&Volume&Normalized Volume&Speedup to achieve $P_{adm}$\\
\hline $P_{adm}$&$4.921\times 10^{-9}$&1&1\\
$P_{coding}$&$4.686\times 10^{-9}$&0.952&1.25\\
$P_{fs}$&$4.613\times 10^{-9}$&0.937&1.25\\
$P_{nofs}$&$2.260\times 10^{-9}$&0.460&1.67\\
\hline
\end{tabular} \end{center}\end{table}

\begin{table}
\begin{center}\caption{A comparison of the four schemes in a $4\times 3$ switch with unicasts and broadcasts only}\label{tab:poly2}
\begin{tabular}{|l|c|p{1.5cm}|p{1.8cm}|}
\hline
Polytope&Volume&Normalized Volume&Speedup to achieve $P_{adm}$\\
\hline $P_{adm}$&$1.4546\times 10^{-9}$&1&1\\
$P_{coding}$&$1.4541\times 10^{-9}$&0.9997&1.25\\
$P_{fs}$&$1.4527\times 10^{-9}$&0.9987&1.25\\
$P_{nofs}$&$1.0585\times 10^{-9}$&0.7277&1.67\\
\hline
\end{tabular} \end{center}\end{table}

In Table \ref{tab:poly} and Table \ref{tab:poly2}, the rate regions are compared in term of the volume of the polytope and the minimum speedup needed to achieve 100\% throughput. Here, $P_{adm}$ refers to the admissible region; $P_{coding}$ is the linear intra-flow  network coding rate region; $P_{fs}$ refers to the rate region with fanout-splitting only; and $P_{nofs}$ is the rate region when fanout-splitting is not allowed.

The methodology we used to compute these values was to list all the stable sets of the enhanced conflict graph using a greedy algorithm. Using these list of stable sets and a MATLAB packet called the multi-parametric toolbox \cite{mpt}, we computed the stable set polytope in terms of linear inequalities which in tern gave us the rate region. Once we have an explicit description of the rate regions, we used a software package known as Vinci \cite{vinci} to compute the volume of the rate regions. The rate region of the case with fanout splitting but no coding was obtained using the characterization given by Marsan \etal \cite{marsan}. The speedup required to achieve $P_{adm}$ is equal to the minimum factor needed to expand the polytope such that it covers $P_{adm}$.

It is interesting to note that the speedup needed to achieve 100\% throughput for $P_{coding}$ and $P_{fs}$ is 1.25 for $2\times 3$ and $3\times 4$ switch. Furthermore, we verified that the traffic patterns that require speedup of 1.25 to achieve are variations of the traffic pattern shown in Figure \ref{fig:5/4}. This seems to indicate that the ``hardest'' admissible traffic patterns to achieve are those that have an enhanced conflict graph with an odd hole of length 5. This observation leads to our Conjecture \ref{thm:5/4} (presented in Section \ref{sec:summaryCodingvsSpeedup}) that the actual minimum speedup required to achieve 100\% throughput in a $K \times N$ switch with traffic patterns consisting of unicasts and broadcasts only is exactly 5/4 when network coding is allowed.

It may seem that the results in Table \ref{tab:poly} and \ref{tab:poly2} show that coding does not outperform fanout-splitting by much in terms of total achievable rate region. However, we should not interpret this result as such.  Another way of looking at the two polytopes $P_{coding}$ and $P_{fs}$ is to just compare these two directly. From our previous example in Figure \ref{fig:benefit}, we know that $P_{fs} \subsetneq P_{coding}$. So, we can ask what is the speedup needed for $P_{fs}$ to achieve $P_{coding}$. For our $2\times 3$ switch and $4\times 3$ switch, the speedup we need for the fanout-splitting region to achieve the network coding region is 1.1667. The traffic patterns that require this speedup are variations of the traffic pattern shown in Figure \ref{fig:benefit1}, which requires a speedup of $(1.5 - \frac{1}{N}) = \frac{7}{6} \approx 1.1667$ (where $N = 3$) when fanout-splitting is allowed as shown in Theorem \ref{thm:examplerateregion}. Therefore, this shows that network coding can give us a benefit equivalent to speedup of at least 1.1667.

Interestingly, the speedup required for $P_{nofs}$ to achieve 100\% throughput is 1.67 in both $2\times 3$ and $4\times 3$ switch, and the traffic patterns that require this speedup are also the variations of the traffic pattern shown in Figure \ref{fig:benefit1}. These observations indicate that there may be a few traffic patterns that are ``hard'' to achieve, and focusing our analysis on these few traffic patterns may be enough to understand the performance of a scheme in general. If this is the case, the challenge lies in finding these few key traffic patterns, and network coding with its graph-theoretic interpretation gives us insights into which traffic patterns might be of significance.

\subsection{A conjecture on the minimum speedup}\label{sec:summaryCodingvsSpeedup}
We have thus introduced a simple graph-theoretic bound on the speedup needed to achieve 100\% throughput in a multicast network coding switch using the concept of conflict graphs. We have shown that the imperfection ratio of the enhanced conflict graph gives an upper bound on the speedup needed.

Applying this result to the special case of $K\times N$ switches with unicasts and broadcasts only, we have obtained an upper bound on speedup of $\min (\frac{2K-1}{K}, \frac{2N}{N+1})$.
For a $2\times N$ switch, the upper bound evaluates to 1.5. We showed earlier in this section that the speedup is lower bounded by 1.25 for any switch with 2 or more inputs and 3 or more outputs. We have verified using a computer that the actual speedup needed is 1.25 for the case of $2\times 4$, $2\times 5$, $3\times 3$ and $4\times 3$ switches, which meets the lower bound. This seems to indicate that the enhanced conflict graph for the case of unicasts and broadcasts has a structure that fixes the imperfection ratio at 1.25. We will next present some results and conjectures on this structure.

Consider a $2\times N$ switch. We use the same notation as in Section \ref{sec:KxN}. Let $I$ and $O$ denote the set of inputs and outputs respectively. Let $G$ denote the enhanced conflict graph. Then, the fractional stable set polytope $QSTAB(G)$ is given by the following inequalities:
\begin{eqnarray}
\label{u1j}\mbox{For }j\in [N], \ \ \ \ u_{1j}&\ge& 0\\
\label{b1j}\mbox{For }j\in [N], \ \ \ \ b_{1j}&\ge& 0\\
\label{u2j}\mbox{For }j\in [N], \ \ \ \ u_{2j}&\ge& 0\\
	\label{ip1ineq}\mbox{For } j\in [N],\ \ \ \ b_{1j}+\sum_{k=1}^nu_{1k}&\le &1\\
	\label{ip2ineq}\sum_{j=1}^n u_{2j}&\le & 1\\
	\label{opineq}\mbox{For } j\in [N],\ \ \ \ u_{1j}+u_{2j}+b_{1j}&\le &1
\end{eqnarray}

Every stable set of $G$ satisfies the clique conditions and is therefore a part of $QSTAB(G)$. Now, $QSTAB(G)$ is clearly inside the unit hypercube $[0,1]^{3N}$. Therefore, since the stable sets are 0-1 vectors, they cannot be expressed as a non-trivial convex combination of two distinct points in $QSTAB(G)$. This means every stable set of $G$ is an extreme point of $QSTAB(G)$. The following theorem specifies some other extreme points.
\begin{theorem}\label{thm:cornerpts}\it
The vectors $\mathbf{v}(m,U,V)$ of the following form are extreme points of $QSTAB(G)$:
\begin{eqnarray*}
 u_{1m}&=&|U|^{-1} \\
 b_{1j}&=&1-|U|^{-1} \mbox{ for all }j\in V\\
 u_{2j}&=&|U|^{-1} \mbox{\ \ \  for all }j\in U
\end{eqnarray*}
where $U$ runs over all subsets of $O$ such that $2\le |U|\le (n-1)$, and for a given $U$, $m$ runs over all outputs in $O\backslash U$ and $V$ runs over all subsets of $O$ such that $V\supseteq U$. The rates of all other subflows are zero. (See Figure \ref{fig:cornerpoint}).
\end{theorem}
The proof is given in the appendix. Note that although the figure shows a case where $m\notin V$, in general, $m$ could be in $V$.

\begin{figure}[h!]
\begin{center}
\includegraphics[width=0.35\textwidth]{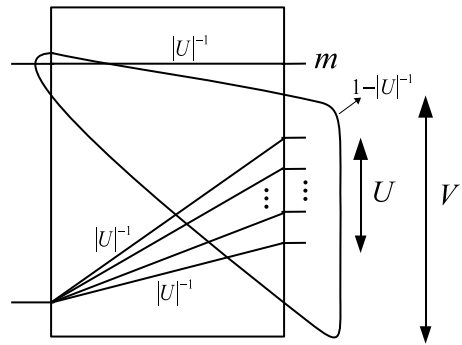}
\end{center}
\caption{Extreme point of $QSTAB(G)$}\label{fig:cornerpoint}
\end{figure}

\begin{theorem}\label{thm:fracchrom}\it
The fractional weighted chromatic number of $G$ with the weight vector set equal to the point $\mathbf{v}(m,U,V)$ from Theorem \ref{thm:cornerpts} is upper bounded by $1+|U|^{-1}-|U|^{-2}$ and hence is no larger than 1.25.
\end{theorem}
The proof is given in the appendix.

\begin{conjecture}
$QSTAB(G)$ has no other extreme points besides the stable sets and the ones given in Theorem \ref{thm:cornerpts}.
\end{conjecture}

If this conjecture is true, then Theorem \ref{thm:fracchrom} will imply that an expansion factor of 1.25 will enable $STAB(G)$ to cover every vertex of $QSTAB(G)$ for a $2\times N$ switch with unicasts and broadcasts only. Based on our simulations, we believe that this approach will in fact extend to a $K\times N$ switch with unicasts and broadcasts. This leads to the following conjecture.

\begin{conjecture}\label{thm:5/4}
\it The minimum speedup required to achieve 100\% throughput in a $K\times N$ switch with traffic patterns consisting of unicasts and broadcasts only is exactly 1.25.
\end{conjecture}

If true, this conjecture shows that the ``worst''  traffic pattern induces an enhanced conflict graph that contains an odd hole of size 5 (for example, Figure \ref{fig:5/4}).

In summary, by allowing network coding in multicast switches, we derive not only a graph-theoretic characterization of the speedup needed for 100\% throughput, but also a gain in throughput and speedup. We have shown that network coding, which is usually implemented using software, can substitute speedup, which is often achieved by adding extra switch fabrics.

\section{Algorithms for offline and online scheduling}\label{sec:algorithms}
In this section, we propose offline and online scheduling algorithms to achieve the rate region of network coding multicast switches. We first start with the offline algorithm in Section \ref{sec:offline}, then discuss the maximum weighted stable set (MWSS) online algorithm in Section \ref{sec:online} and its refinement in Section \ref{sec:finitehorizon}. In Section \ref{sec:sim_improvdelay}, we study the effect of network coding in an online setting via simulations.

\subsection{Rate decomposition approach for offline scheduling}\label{sec:offline}
The proof of Theorem \ref{thm:frame} gave an offline scheduling strategy which used the fact that as long as the enhanced rate vector is within the stable set polytope of the enhanced conflict graph, it can be decomposed into a convex combination of valid switch configurations. Thus, prior knowledge of the average arrival rates of the flows can be used to obtain a schedule. In this subsection, we focus on this problem of rate decomposition for offline computation of the schedule in a manner similar to the Birkhoff-von Neumann switch for unicast \cite{bvnswitches}. The following discussion gives a graph-theoretic interpretation of this problem.

Recall that the fractional weighted coloring problem involves decomposing a vector of weights on the vertices of a graph into a linear combination of stable sets (see Section \ref{sec:graphtheory}). We interpret the weights to correspond to the flow rates, and the coefficients $\lambda_i$ used in the linear combination to be the fractions of time in the schedule. If the fractional weighted chromatic number is less than 1, then the optimal solution expresses the weight vector as a convex combination of stable sets, which in turn leads to a switch schedule. This leads to the following theorem.

\begin{theorem}
\it The problem of computing the offline switch schedule for a multicast traffic pattern when fanout splitting and intra-flow linear network coding are allowed, is equivalent to the problem of fractional weighted coloring of the enhanced conflict graph, with the enhanced rates used as vertex weights.
\end{theorem}

If the fractional weighted chromatic number $c$ for a given rate vector exceeds 1, then such a rate vector cannot be achieved, since it is not within the stable set polytope. However, if the rate vector is admissible, then it can be achieved if we allow a speedup equal to $c$. This leads to an interesting physical interpretation for the fractional weighted chromatic number corresponding to a given admissible rate vector, summarized in the following theorem.
\begin{theorem}
\it The minimum speedup needed to achieve an admissible rate vector with fanout splitting and coding, is the fractional weighted chromatic number of the enhanced conflict graph, with the enhanced rate vector used as vertex weights.
\end{theorem}
\IEEEproof
Let $c$ be the fractional weighted chromatic number of a given rate vector. A speedup of $s$ means that the switch can go through $s$ configurations per time-slot. Thus, from the point of view of the input queues, their service rate is now $s$ times higher. Equivalently, if we redefine a slot to be the time spent by the switch in each configuration, then the speedup essentially scales down the arrival rate vector seen by the input queues by a factor of $s$. This means, for a given rate vector, the input queues can be stabilized with speedup $s$ if the same rate vector, when scaled down by a factor of $s$, is achievable without any speedup. However, scaling down the rate vector by $s$ will also scale down the optimum value of the fractional weighted coloring problem by the same factor, \ie, it will now be $c/s$. Thus, if $c\le s$, then the new scaled rate vector is achievable without speedup. Therefore, for the given traffic pattern, the input queues can be stabilized with the speedup.

But this is only from the point of view of the input queues. Note that with speedup, there are queues at the outputs of the switch as well and achievability means stabilizing both the input and output queues. Now, stability of the output queues only requires that the net inflow into an output queue must not exceed 1, which is part of the admissibility conditions. Since the traffic pattern is given to be admissible, we get the required result.
\endproof

This switch schedule and the network code which ensures that every transmission conveys an innovative packet, together give a complete specification of a frame-based scheme that achieves the entire rate region, as shown in Section \ref{sec:achievablerateregion}.

\subsection{Maximum weighted stable set algorithm for online scheduling}\label{sec:online}
Suppose the rates of the various flows are unknown and scheduling has to be done online using only the current queue occupancy information. Analogous to the maximum weighted matching algorithm for unicast \cite{100forunicast}, we show that for multicast switches with fanout splitting and network coding, a maximum weighted stable set (MWSS) algorithm on the enhanced conflict graph achieves the same rate region as is achievable with prior knowledge of rates. In this section, we assume that the arrivals to each flow are \emph{i.i.d.} and independent across flows. In this section, we first examine the conditions under which the virtual queues can be served in a stable manner. In the next section, we will show that this stability can also be extended to the physical queues.

Let $q_{iJj}(t)$ denote the occupancy of the virtual queue for subflow $(i,J,j)$ in time-slot $t$. Thus, $q_{iJj}(t)$ is a measure of the backlog for subflow $(i,J,j)$ in terms of the degrees of freedom. The MWSS algorithm uses these virtual queue sizes as weights to compute the maximum weighted stable set. We now present the algorithm.

\begin{table}\label{maxwtstableset}
\centering
\emph{Algorithm:} Max Weighted Stable Set (MWSS)\\
\ \\
\begin{tabular}{lp{7.5cm}}
1. & Using $q_{iJj}(t)$ as the weight for the vertex corresponding to the subflow $(i,J,j)$, compute the maximum weighted stable set in the enhanced conflict graph. This specifies the set of subflows that will be served in the current time-slot. If $q_{iJj}$ is 0 for any chosen subflow, it is dropped from the set. \\
2. & For every flow in the chosen set, compute a linear combination of all packets received for that flow until time $t$, such that, the linear combination is innovative for all the chosen outputs of that flow. (It will be proved below that this is always possible.) \\
3. & Transfer the computed linear combination to the outputs of the subflows chosen in the stable set in step 1, and update $q_{iJj}(t)$ accordingly. Go back to step 1.
\end{tabular}
\vspace{-.2in}
\end{table}

\begin{lemma}\label{subspacelemma}
\it Let $\mathcal{V}$ be a vector space with dimension $n$ over a field of size $q$, and let $\mathcal{V}_1, \mathcal{V}_2, \ldots \mathcal{V}_k$, be subspaces of $\mathcal{V}$, of dimensions $n_1, n_2, \ldots, n_k$ respectively. Suppose that $n>n_i$ for all $i=1, 2, \ldots, k$. Then, there exists a vector that is in $\mathcal{V}$ but is not in any of the $\mathcal{V}_i$'s, if $q>k$.
\end{lemma}
\IEEEproof
The total number of vectors in $\mathcal{V}$ is $q^n$. The number of vectors in $\mathcal{V}_i$ is $q^{n_i}$. Hence, the number of vectors in $\cup_{i=1}^k\mathcal{V}_i$ is at most $\sum_{i=1}^k q^{n_i}$.
Now,
{\begin{center} $\sum_{i=1}^k q^{n_i} \leq kq^{n_{max}} \leq kq^{n-1} < q^n$\end{center}}
\noindent where, $n_{max}$ is $\max_i n_i$, which is at most $(n-1)$.
Thus, $\mathcal{V}$ has more vectors than $\cup_{i=1}^{k}\mathcal{V}_i$. This completes the proof.
\endproof

\begin{remark}
For the above algorithm to work, we need to show that in the second step, there is at least one linear combination which is guaranteed to be innovative to all chosen outputs. Now, $q_{iJj}$ gives the difference between the dimension of the knowledge space of input $i$ and that of the output $j$ for flow $(i,J)$. Hence, if $q_{iJj}$ is positive for a set of outputs, then we have the same situation as in Lemma \ref{subspacelemma}. The $k$ subspaces in the lemma correspond to the knowledge spaces of the outputs, while $n$ is the dimension of the overall knowledge space of the input. Thus, the lemma guarantees that there exists a linear combination of the packets of flow $(i,J)$ that is innovative to all those outputs, as long as the field size is larger than the number of outputs involved. Such a combination is chosen in step 2.

Note that while this argument shows the existence of such a linear combination, it does not give an explicit way to find one. However, since the scheduling and coding is done in a centralized manner, the encoder knows the outputs' knowledge spaces completely. Hence, the algorithm proposed in \cite{ARQforNC} can be used to compute the required linear combination.
\end{remark}

\begin{theorem}\label{mwss}
\it If the arrivals are i.i.d. and independent across flows and the rate vector is inside $\mathbf{R}$ (the rate region in Corollary \ref{thm:rateregion}), then the MWSS algorithm given above, stabilizes the virtual queue size vector $\mathbf{q}$ in the mean.
\end{theorem}
\IEEEproof
The proof is essentially an application of the results of~\cite{tassephrem} and~\cite{tassiulas} for the case of parallel queues. Consider the virtual queues as a system of parallel queues.

It is clear that two virtual queues which conflict with each other cannot be served at the same time. Lemma \ref{subspacelemma} and the remark above imply that the converse is also true, \ie, any set of non-empty virtual queues which have no conflicts \emph{can} be served simultaneously. The virtual queues corresponding to the chosen stable set will all receive one unit of service, since the reception of an innovative packet by the output will decrease the difference in dimension of the knowledge space between the input and output by 1. Using the terminology in \cite{tassephrem}, this means that \emph{eligible activation vectors} of the queues therefore correspond to conflict-free sets of subflows, or in other words, stable sets in the enhanced conflict graph.

The only difference between this situation and the one assumed in~\cite{tassephrem} is that~\cite{tassephrem} assumes that arrivals to different queues are independent of each other, whereas in our case, arrivals to subflows of the same flow always occur simultaneously. However, this lack of independence across arrival processes does not affect the results of~\cite{tassephrem}, essentially because of the linearity of expectation of dependent random variables. Stability in the mean still holds, as long as other assumptions such as the ergodicity of the arrival processes and the finiteness of their second moment hold. Thus, the MWSS algorithm stabilizes the occupancy of the virtual queues ($\mathbf{q}$), as long as their arrival rates are inside the convex hull of the eligible activation vectors, which is the stable set polytope of the enhanced conflict graph. In other words, as long as the arrival rate vector is within $\mathbf{R}$, $\lim_{t\rightarrow \infty}E[q_{iJj}(t)]<\infty$ $\forall$ subflows $(i,J,j)$.
\endproof

\subsection{Stabilizing the physical queues}\label{sec:finitehorizon}
In this section, we will connect the virtual queue size to the physical queue size. The above theorem shows that the MWSS algorithm stabilizes the backlog in number of degrees of freedom. However, the algorithm does not specify any rule for packets to depart from the physical buffer at the inputs.

We propose a departure rule based on the following observation. If a packet has been decoded by all the outputs in the fanout of the packet's flow, then involving such a packet in the transmitted linear combination does not convey anything new that could not have been conveyed without involving this packet. This naturally implies the following departure rule -- a packet departs from the physical queue when it has been decoded by all outputs in the fanout of the packet's flow.

To understand the queue dynamics under this departure rule, we need to specify the decoding mechanism. We assume a centralized system where the output knows the coefficients used by the input in the linear combinations. The output can verify if a packet is innovative by checking whether its coefficient vector is in the output's knowledge space. Each innovative packet counts as a new degree of freedom and provides a new equation in the packets. With enough degrees of freedom, the output simply needs to invert the matrix of coefficients to obtain the original packets. Thus, if the backlog becomes 0 (\ie, the virtual queue becomes empty), the number of equations becomes equal to the number of unknowns and the output can completely decode all the packets till that point.

Consider a slot in which all virtual queues of a flow become empty simultaneously. At this point, the physical queue for that flow will also be empty, because at this point, all outputs in the flow's fanout will have reached the state of having decoded all packets.

Now, the fact that virtual queues are stable in the mean implies that the underlying Markov chain is positive recurrent. Thus, the chain will visit the state where all virtual queues are empty, infinitely often with probability one. The connection between the emptying of the virtual queues of a flow and the physical queue of that flow implies that the physical queue will also become empty infinitely often almost surely. This conclusion is summarized by the following corollary.

\begin{corollary}\label{cor:positiverecur}
\it If the arrivals are i.i.d. and independent across flows and the mean arrival rate vector is strictly inside the rate region $\Gamma$, then the strategy of allowing a packet to depart when it is decoded by all outputs in its fanout ensures that the physical reaches a the empty state infinitely often, with probability (w.p.) 1.
\end{corollary}
\begin{remark}
The strategy of dropping a packet when it has been decoded by all outputs in its fanout, can be improved using a streaming policy for buffer clearance, as proposed in \cite{ARQforNC}. This work introduces the notion of a node ``seeing'' a packet. Using that notion, packets of a flow that have been seen by all outputs in the flow's fanout can be dropped from the queue. When combined with the coding module proposed in \cite{ARQforNC}, this will allow the physical buffer size to follow the virtual queue sizes without compromising on throughput. Thus, the stability results of the virtual queue readily carry over to the physical queue as well. This approach allows to prove stability of the physical queue in the mean, which is stronger than the positive recurrence claimed in Corollary \ref{cor:positiverecur}.
\end{remark}

\begin{remark}
The results in~\cite{harish} are related to our approach. In that paper, the authors analyze the performance of a back-pressure based policy for wired and wireless networks with intra-flow coding, and show that it stabilizes the system for all rate vectors within the capacity region. The network constraints are captured in terms of capacities on each link, which could be inter-dependent in the wireless setting. The crossbar switch, studied in our paper, is similar to the wireless setting in the sense that, an input may not send a different packet to different outputs simultaneously. Besides, there is a special kind of inter-dependence among the links in that, of all links going to the same output, at most one may be active at a time. However, \cite{harish} gives an indirect characterization of the rate region in terms of certain flow variables, unlike the more explicit graph-theoretic characterization we have provided.
\end{remark}

\subsection{Simulations: Improvement in delay}\label{sec:sim_improvdelay}
In this section, we study how network coding, even if it does not improve the rate, can decrease delay dramatically. We study the effect of coding in an online setting, through MATLAB simulations in a $2\times 4$, $3\times 3$, and $4\times 3$ switches. The setup we use is similar to the MWSS algorithm, which stabilizes the virtual queues as well as the physical queues (Section \ref{sec:online} and Section \ref{sec:finitehorizon}). We modify the MWSS algorithm in two ways for the simulation.

First, we use a batching-version of the MWSS algorithm -- packets are grouped into batches according to their arrival times. The batch length is denoted as $\Delta_0 = \Delta(1+\epsilon)$, and all arrivals from time $k\Delta_0$ to $(k+1)\Delta_0$ are said to belong to batch $k$. The basic idea is to run MWSS on one batch for $\Delta$ slots, then take a break to clear the backlog for that batch during the following $\epsilon \Delta$ slots, thereby allowing the outputs to decode it completely. After that, the batch is flushed out of the input buffers, and then, we begin afresh with the next batch. These breaks will cause a loss in throughput, since the MWSS algorithm is now running for only a fraction of the time. However, with a large enough batch length, this throughput loss can be made arbitrarily small.

Second, instead of the maximum weight stable set which is known to be $NP$-hard~\cite{karp}, we use a simpler randomized algorithm using an idea proposed in \cite{tassiulas}. Reference \cite{tassiulas} proposes a scheduling approach that leads to policies with maximum throughput and yet linear complexity per packet transmission for a resource allocation problem for several computer and communication network architecture. The proposed policy is a randomized, iterative algorithm with a combination with an incremental updating rule. Although there is no guarantee that at any time the configuration used is  optimal, the policy approximates the optimal policy such that it provides maximum throughput.

Our randomized algorithm approximates the MWSS algorithm. In each slot, we randomly generate a constant number of maximal stable sets. Given the current backlog, we compute the weight of all the randomly generated maximal stable sets as well as the stable set that was used in the previous slot. Then, we select the maximum weight stable set and use it as the configuration of the current slot.

We compare the performance with the case of fanout splitting without coding. For this case, we use a similar randomized modification of the algorithm given in~\cite{marsan}. Instead of stable sets, we use the modified departure vectors defined in~\cite{marsan}.

\begin{figure}[tbp]
\begin{center}
\includegraphics[width=0.48\textwidth]{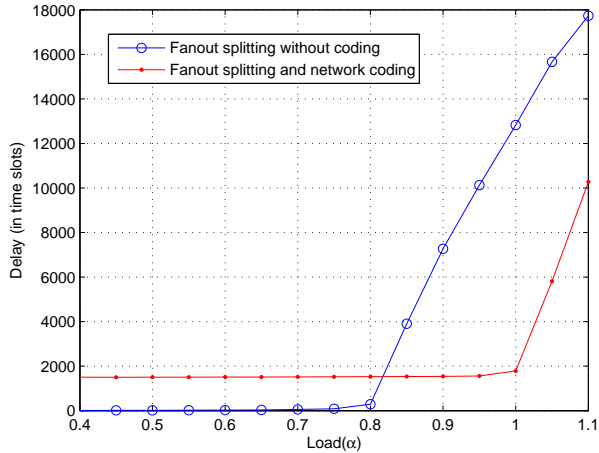}
\end{center}
\caption{Delay vs. load plots with and without coding in a $2\times 4$ switch}
\label{fig:24delay}
\end{figure}

In Figure \ref{fig:24delay}, we study the performance of a $2\times 4$ switch with and without coding. For this simulation, the parameters $\Delta$ and $\epsilon$ in the finite horizon MWSS algorithm were set to be 3000 and 0.005. The traffic pattern used here is identical to that of in Figure \ref{fig:benefit1} with $N = 4$. Arrivals are generated according to an {\it i.i.d.} Bernoulli process independently for each flow.

In Section \ref{sec:sim_improverate}, we discussed that the traffic pattern in Figure \ref{fig:benefit1} is achievable when network coding is allowed while it is not if we only allow fanout-splitting, which is reflected in the plot in Figure \ref{fig:24delay}. At light loads, the algorithm using coding incurs a larger delay due to coding and decoding costs. When the traffic is light, inputs of the uncoded scheme just relay the packets to the outputs; however, in the coded scheme, outputs need to wait until they have received enough packets to decode the entire batch. As a result, we see that there is a consistent delay of approximately 1500 slots for the coded scheme at light loads. It is important to note that the delay of 1500 slots is not an arbitrary delay, but a parameter we can choose depending on our application. The delay is the average slots each packet has to wait until it is decoded at the output -- and since our batch size $\Delta$ is 3000, the average delay is 1500.

The interesting part of our result is when the load is heavier. First we note that, for the uncoded scheme, the delay increases dramatically at a lower value of load ($\alpha \approx 0.8)$, as opposed to the coded scheme ($\alpha \approx 1$). Thus, in terms of throughput, the coded scheme is better. This empirically shows that network coding increases the rate region. Here, we note the significance of the two boundary values: $\alpha \approx 0.8$ and $\alpha \approx 1$. First, as mentioned above, the traffic pattern in Figure \ref{fig:benefit1} is achievable with coding; thus, as we expect, coding does not incur heavy delays until $\alpha \approx 1$. Second, Theorem \ref{thm:examplerateregion} in Section \ref{sec:sim_improverate} prove that a speedup of at least $(1.5 - \frac{1}{N})$ is needed to achieve 100\% throughput with fanout-splitting only. In this example, we have $N = 4$; therefore, we need speedup of at least $1.5 - \frac{1}{4} = \frac{5}{4}$, which is reciprocal of $\alpha \approx 0.8$.

\begin{figure}[tbp]
\begin{center}
\includegraphics[width=0.48\textwidth]{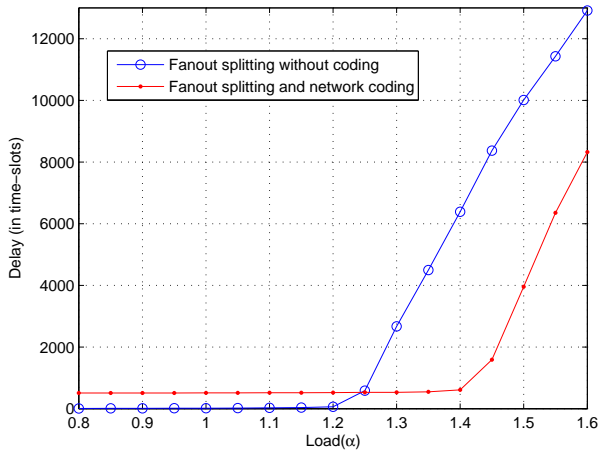}
\end{center}
\caption{Delay vs. load plots with and without coding in a $3\times 3$ switch}
\label{fig:33delay}
\end{figure}

\begin{figure}[tbp]
\begin{center}
\includegraphics[width=0.48\textwidth]{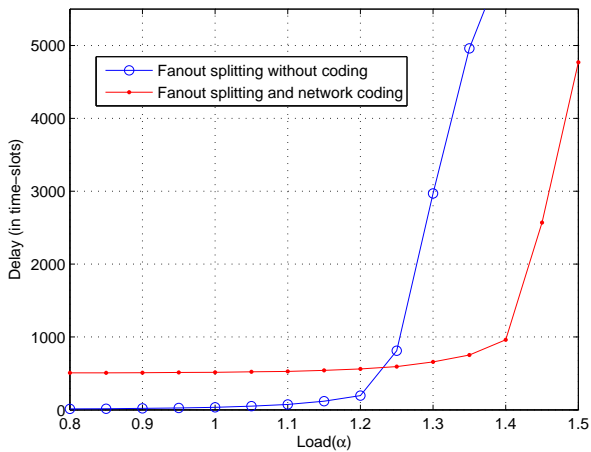}
\end{center}
\caption{Delay vs. load plots with and without coding in a $4\times 3$ switch}
\label{fig:delay}
\end{figure}

In Figure \ref{fig:33delay} and Figure \ref{fig:delay}, we study the performance of a $3\times 3$ and a $4\times 3$ switch with and without coding. For this simulation, the parameters $\Delta$ and $\epsilon$ in the finite horizon MWSS algorithm were set to be 1000 and 0.005. In these two simulations, we use a more general traffic pattern which is a combination of the example pattern in Figure \ref{fig:codinggood} weighted by a factor of $\frac{2}{3} \alpha$, and a pattern with all uniform unicasts, each having a rate of $0.01\alpha$, where $\alpha$ represents the load factor. Therefore, the traffic pattern for Figure \ref{fig:33delay} consists of one broadcast from input 1, with a rate of $\frac{4}{9} \alpha$. There are three unicasts, one to each output, from inputs 1 and 3, each having a rate of $0.01\alpha$. From input 2, there is a unicast of rate $(\frac 29+0.01)\alpha$. The traffic pattern for  Figure \ref{fig:delay} is identical to that of Figure \ref{fig:33delay} with additional three unicasts, one to each output, from input 4 with a rate of $0.01\alpha$ each.

Figure \ref{fig:33delay} and Figure~\ref{fig:delay} show the plot of delay vs. load for the randomized algorithm with and without coding in a $3\times 3$ and a $4\times 3$ switch. As mentioned above, at light loads, the algorithm using coding incurs a larger delay of approximately 500 due to coding and decoding costs, which is consistent with the parameters we have chosen ($\Delta = 1000$).

Again, network coding shows its strength when the load is heavier. In both simulations in Figure \ref{fig:33delay} and Figure \ref{fig:delay}, we see an increase in throughput. The difference in $\alpha$ in which the delay increases dramatically for coding and fanout-splitting is approximately 0.2 for both simulations. This empirically shows that network coding increases the rate region. Equivalently, network coding leads to delay benefits at high loads. We can consider the load beyond $\alpha = 1.4$ in Figure \ref{fig:delay} for instance. Here, the traffic load is outside of the rate region for with and without network coding. Therefore, we would expect the delay for both coded and uncoded schemes to surge up. The part that interests us is the significant difference in delay between the two schemes. This shows that under heavy traffic, network coding is robust and, although the traffic pattern is beyond its rate region, it delivers the packets with much smaller delay than the uncoded scheme even when we take the coding and decoding cost into account.

\section{Conclusion}\label{sec:conclusion}
In this paper, we explore some issues regarding the benefit of network coding in multicast switch in terms of throughput, delay, and  speedup. Although network coding includes coding schemes with any arbitrary functions, we focus our attention to linear network coding.  This is because linear network coding is sufficient to achieve capacity in a multicast switch, and it gives us simplicity in code. We show that allowing linear intra-flow network coding at the inputs leads to a larger rate region in general. We demonstrate examples of traffic patterns  where coding eliminates the need for speedup to serve the traffic in a stable manner. In addition, using linear network coding allows us to use a graph-theoretic formulation called the conflict graph from \cite{conflictgraphs}, which is an insightful formulation that brings the problem  to its combinatorial essence.

In summary, by allowing network coding in multicast switches, we get not only a characterization of the speedup needed for 100\% throughput, but also a gain in throughput, delay, and speedup. We have shown that network coding, which is usually implemented using software, can substitute speedup, which is often achieved by adding extra switch fabrics. This paper presents a graph-theoretic approach to quantify the minimum speedup needed to achieve 100\% throughput. This new formulation helps us better understand the problem  and enables us to use combinatorial and graph-theoretic results to measure the benefit of network coding in switches.

Possible future work could be to use this formulation to come up with approximation schemes and heuristics that simplify the online  scheduling algorithm and make it practical. Furthermore, studying the benefit of inter-flow coding, which was mentioned briefly in Section \ref{sec:intraflowcoding}, using a similar graph-theoretic approach  could lead to interesting results.

\section*{Acknowledgment}
The authors would like to thank Prof. Devavrat Shah for helpful discussions about this topic.

\appendix
Proof of Theorem \ref{thm:examplerateregion}
\IEEEproof
We will need to show that these inequalities are necessary and sufficient for a rate point in the rate region.

\emph{Necessity: } Equation \ref{eq:nonneg} is the non-negativity constraint; Equation \ref{eq:input2} and Equation \ref{eq:outputs} represent the admissibility conditions for input 2 and each output $i$. Therefore, these equations are necessary conditions for the achievable rate region. We now need to show that Equation \ref{eq:fanout} is also necessary.

Consider any point $\mathbf{r}$ inside the rate region. There is a frame-based schedule with frame size $F$ such that, if input 1 has $r_0F$ packets for the broadcast flow and input 2 has $r_iF$ packets for the unicast flow to output $i$ at the beginning of a frame, then by the end of that frame, both inputs will be able to deliver all these packets to the corresponding outputs. ({\it Note:} In this proof, we assume without loss of generality that all rates are rational numbers, and that $F$ is a large enough integer such that $r_0F$, $r_iF$ etc. are all integers.)

Let $x$ be the number of slots in the schedule in which input 2 is not transmitting. Input 1 can deliver $x$ packets of the broadcast flow to all outputs in this time. In the remaining $(F-x)$ slots, input 1 requires at least two slots to deliver one packet to all outputs, since in any slot, at least one of the outputs is blocked by input 2. Hence, the number of packets delivered in this time is at most $\frac{(F-x)}{2}$. Therefore, in order to satisfy the requirement, we need:
\[x+\frac{(F-x)}2 \ge r_0F\]
As for input 2, it has $(F-x)$ slots to process all the unicast packets. Thus, to satisfy the unicasts, we need:
\[F-x\ge \left(\sum_{i=1}^N r_i\right)F\]
Eliminating $x$ between the above two conditions and canceling $F$ throughout gives Equation \ref{eq:fanout}.

\emph{Sufficiency: } To show that Equations \ref{eq:nonneg} through \ref{eq:fanout} are sufficient, we will provide an explicit construction for a frame-based schedule that achieves any rate point $\mathbf{r}$ inside the given polytope. We will show that if we start with a collection of $r_iF$ packets for flow $i$, then all these packets can be correctly delivered to their destined outputs within $F$ slots. For convenience, we denote $S:=\sum_{i=1}^N r_i$. Define $\alpha:=\min\left(\frac{r_0}S, \frac12\right)$.

Partition the $r_0F$ packets of the broadcast flow into $N+1$ groups in the following way. Place the first $\alpha r_1F$ packets in the group 1. Place the next $\alpha r_2F$ packets in group 2, and so on. After filling group $N$ with $\alpha r_NF$ packets, place the remaining packets (if any) in group $(N+1)$. ({\it Note:} We can choose a large enough $F$ such that $\alpha r_iF$ is an integer for all $i$.)

\begin{figure}[h!]
\begin{center}
\includegraphics[width=0.5\textwidth]{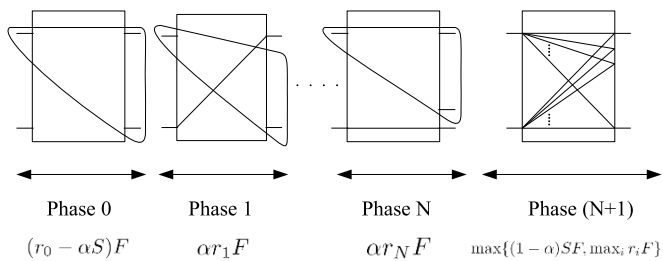}
\end{center}
\caption{The schedule for proof of Theorem \ref{thm:examplerateregion}}\label{fig:phases}
\end{figure}

The schedule consists of $N+2$ phases, numbered 0 to $N+1$, as shown in Figure \ref{fig:phases}. In phase 0, serve packets in group $N+1$ (if any) using a broadcast connection to all outputs. This requires $(r_0-\alpha S)F$ slots.

In phase 1 which lasts for the next $\alpha r_1F$ slots, serve $\alpha$ fraction of the unicast from input 2 to output 1. Simultaneously transmit all group 1 packets from input 1 to all outputs except output 1. Similarly, in phase 2 (the next $\alpha r_2F$ slots), serve $\alpha$ fraction of the unicast to output 2 along with broadcasting group 2 to other outputs, and so on.  Phases 1 to $N$ require $\alpha SF$ slots to complete.

At the end of phase $N$, exactly $\alpha$ fraction of all unicasts have been served. In addition, each output $i$ has also received the broadcast flow packets from all groups except group $i$. But this means that each broadcast flow packet now needs to reach only one output. Thus, we essentially have a unicast problem, since no two outputs want the same packet. These ``unicasts'' are served in phase $N+1$ along with the unicasts from input 2. We now need to bound the duration of this phase. From the unicast switch literature, it is well known that the minimum number of slots to serve out all the unicasts is equal to the maximum load in terms of number of packets at any of the input or output ports. This follows from a theorem of Birkhoff \cite{birkhoff}, and is stated in Fact 1 of \cite{integrated}.

The load on input 1 is simply the sum of the sizes of the first $N$ groups, which is $\alpha SF$. The load on input 2 from the remainder of the unicasts is $(1-\alpha) SF$ packets. Since $\alpha\le \frac12$, input 2's load dominates input 1's load. Consider output $i$. The load on this output from input 1 is the size of group $i$, which is $\alpha r_iF$. From input 2, it is $(1-\alpha)r_iF$. Thus, the total load on output $i$ is $r_iF$. This means, the duration of phase $N+1$ is $\max\{(1-\alpha) SF, \max_i r_iF\}$.

Summing over all the phases, the total duration of the schedule is: $r_0F+\max\{(1-\alpha) SF, \max_i r_iF\}$.

If $\frac{r_0}{S}\le \frac12$, this gives $r_0F + \max\{(S-r_0)F, \max_i r_iF\}$. From Equations \ref{eq:input2} and \ref{eq:outputs}, this is at most $F$.

If $\frac{r_0}{S}>\frac12$, the duration is $r_0F+\max\{\frac{SF}2, \max_i r_iF\}$. From Equations \ref{eq:outputs} and \ref{eq:fanout}, this is at most $F$.

Thus, we have presented a schedule which serves all the packets within $F$ slots.
\endproof

\

Proof of Theorem \ref{thm:cornerpts}
\IEEEproof
We make use of the following fact about polytopes (see Theorem 5.7 in \cite{sch}). A vector $\mathbf{z}$ is an extreme point of a polytope of the form $P=\{\mathbf{x}\in \mathbb{R}^n|A\mathbf{x}\le \mathbf{b}\}$ if and only if the sub-matrix of $A$ obtained by including only those rows of $A$ corresponding to constraints that are tight at $\mathbf{z}$, has a rank of $n$.

Now, $QSTAB(G)\subset \mathbb{R}^{3N}$. This means, for the vector in the theorem statement, we need to find $3N$ linearly independent constraints that are satisfied with equality, for every allowed choice of $U$, $V$ and $m$.

Choose $U$, $V$ and $m$ in any way subject to the restrictions in the theorem statement. Consider the following constraints of $QSTAB(G)$: (\ref{u1j}) for $j\neq m$, (\ref{b1j}) for $j\notin V$, (\ref{u2j}) for $j\notin U$, (\ref{ip1ineq}) for $j\in V$, (\ref{opineq}) for $j\in U$ and (\ref{ip2ineq}). There are $3N$ constraints in this list. We set all of them to equality and try to solve the resulting system of equations:

\begin{eqnarray}
\label{eq_u1j}\mbox{For }j\neq m, \ \ \ \ u_{1j}&=& 0\\
\label{eq_b1j}\mbox{For }j\notin V, \ \ \ \ b_{1j}&=& 0\\
\label{eq_u2j}\mbox{For }j\notin U, \ \ \ \ u_{2j}&=& 0\\
	\label{eq_ip1ineq}\mbox{For } j\in V,\ \ \ \ b_{1j}+\sum_{k=1}^nu_{1k}&= &1\\
	\label{eq_ip2ineq}\sum_{j=1}^n u_{2j}&=& 1\\
	\label{eq_opineq}\mbox{For } j\in U,\ \ \ \ u_{1j}+u_{2j}+b_{1j}&=&1
\end{eqnarray}

Now, (\ref{eq_ip1ineq}) implies that the $b_{1j}$'s are all equal for $j\in V$. Let the common value be $b$. Using (\ref{eq_u1j}), (\ref{eq_b1j}) and (\ref{eq_u2j}), and the facts $m\notin U$ and $U\subseteq V$, the last three sets of equations can be simplified to:
\begin{eqnarray}
	\label{eq2_ip1ineq}\ \ \ \ b+u_{1m}&= &1\\
	\label{eq2_ip2ineq}\sum_{j\in U} u_{2j}&=& 1\\
	\label{eq2_opineq}\mbox{For } j\in U,\ \ \ \ u_{2j}+b&=&1
\end{eqnarray}

Now, it is easily seen that this system of equations has a unique solution, and this solution is precisely the rate point $\mathbf{v}(m,U,V)$ given in the theorem statement.

Thus, we have produced $3N$ constraints that are tight at the given point. The fact that the solution is unique implies that the $3N$ constraints considered are linearly independent. This completes the proof.
\endproof

\

Proof of Theorem \ref{thm:fracchrom}
\IEEEproof
To prove this result, we express the weight vector as a linear combination of stable sets $\{S_i\}$ of $G$: $\sum_{i=1}^k\lambda_i\mathbf{\chi}^{S_i}=\mathbf{v}(m,U,V)$ such that $\sum_{i=1}^k\lambda_i=1+|U|^{-1}-|U|^{-2}$ ($\mathbf{\chi}^S$ denotes the incidence vector of the stable set $S$).

\begin{figure}[h!]
\begin{center}
\includegraphics[width=0.5\textwidth]{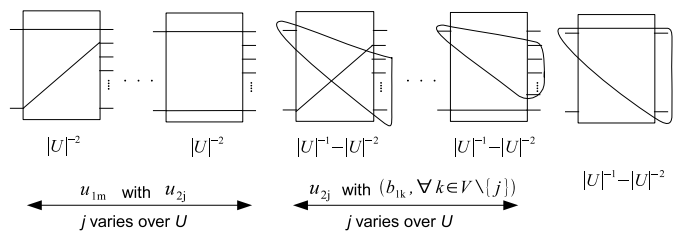}
\end{center}
\caption{The schedule corresponding to Table \ref{tab:schedule}}\label{fig:vertexschedule}
\end{figure}

The stable set and the corresponding $\lambda$ is shown in Table \ref{tab:schedule}. An example of this collection of stable sets and the associated weights is shown in Figure \ref{fig:vertexschedule} in the form of a switch schedule. The figure corresponds to the case where $m=1$ and $U=V=\{2, \ldots N\}$.

It is easily seen that each set given in the table is indeed a stable set. Moreover, the sum of the coefficients $\lambda_i$ is indeed $1+|U|^{-1}-|U|^{-2}$. And finally, the linear combination of the stable sets with the prescribed coefficients gives the rate vector $\mathbf{v}(m,U,V)$, thus completing the proof.
\endproof

\begin{table}
\centering
\caption{The decomposition of the vector in Theorem \ref{thm:cornerpts} into stable sets}\label{tab:schedule}
\begin{tabular}{|c|p{4cm}|p{2.3cm}|}
\hline
&Stable set ($S_i$)& $\lambda_i$\\
\hline
1&For each $j\in U$: $u_{1m}$, $u_{2j}$&$|U|^{-2}$\\
2&For each $j\in U$: $u_{2j}$ with $b_{1k}$ for all $k\in V\backslash\{j\}$ &$|U|^{-1}-|U|^{-2}$\\
3&$b_{1k}$ for $k\in V$&$|U|^{-1}-|U|^{-2}$\\
\hline
\end{tabular}
\end{table}

\bibliographystyle{IEEEtranS}
\bibliography{References}

\begin{thebibliography}{10}
\providecommand{\url}[1]{#1}
\csname url@samestyle\endcsname
\providecommand{\newblock}{\relax}
\providecommand{\bibinfo}[2]{#2}
\providecommand{\BIBentrySTDinterwordspacing}{\spaceskip=0pt\relax}
\providecommand{\BIBentryALTinterwordstretchfactor}{4}
\providecommand{\BIBentryALTinterwordspacing}{\spaceskip=\fontdimen2\font plus
\BIBentryALTinterwordstretchfactor\fontdimen3\font minus
  \fontdimen4\font\relax}
\providecommand{\BIBforeignlanguage}[2]{{%
\expandafter\ifx\csname l@#1\endcsname\relax
\typeout{** WARNING: IEEEtranS.bst: No hyphenation pattern has been}%
\typeout{** loaded for the language `#1'. Using the pattern for}%
\typeout{** the default language instead.}%
\else
\language=\csname l@#1\endcsname
\fi
#2}}
\providecommand{\BIBdecl}{\relax}
\BIBdecl

\bibitem{vinci}
``Vinci - computing volumes of convex polytopes,''
  \emph{http://www.lix.polytechnique.fr/Labo/Andreas.Enge/Vinci.html}.

\bibitem{ahlswede}
R.~Ahlswede, N.~Cai, S.-Y.~R. Li, and R.~W. Yeung, ``Network information
  flow,'' \emph{IEEE Transactions on Information Theory}, vol.~46, pp.
  1204--1216, 2000.

\bibitem{integrated}
M.~Andrews, S.~Khanna, and K.~Kumaran, ``Integrated scheduling of unicast and
  multicast traffic in an input-queued switch,'' in \emph{Proceedings of IEEE
  INFOCOM}, March 1999, pp. 1144--1151.

\bibitem{birkhoff}
G.~Birkhoff, ``Tres observaciones sobre el algebra lineal,'' \emph{Univ. Nac.
  Tucum\`{a}n Revista A}, vol.~5, p. 147.

\bibitem{caramanis}
C.~Caramanis, M.~Rosenblum, M.~X. Goemans, and V.~Tarokh, ``Scheduling
  algorithms for providing flexible, rate-based, quality of service guarantees
  for packet-switching in banyan networks,'' in \emph{Proceedings of the
  Conference on Information Sciences and Systems}, 2004, pp. 160--166.

\bibitem{bvnswitches}
C.-S. Chang, D.-S. Lee, and Y.-S. Jou, ``Load balanced birkhoff-von neumann
  switches,'' in \emph{Proceedings of IEEE Workshop on High Performance
  Switching and Routing}, 2001, pp. 276--280.

\bibitem{chvatal}
V.~Chv\'{a}tal, ``On certain polytopes associated with graphs,'' \emph{Journal
  of Combinatorial Theory}, vol.~18, pp. 138–--154, 1975.

\bibitem{insufficiency}
R.~Dougherty, C.~Freiling, and K.~Zeger, ``Insufficiency of linear coding in
  network information flow,'' \emph{IEEE Transaction on Information Theory},
  vol.~51, pp. 2745--2759, 2005.

\bibitem{interflow}
A.~Eryilmaz and D.~Lun, ``Control for inter-session network coding,'' in
  \emph{Proceedings of NetCod}, January 2007.

\bibitem{neely}
L.~Georgiadis, M.~J. Neely, and L.~Tassiulas, ``Resource allocation and
  cross-layer control in wireless networks,'' \emph{Foundations and Trends in
  Networking}, vol.~1, no.~1, pp. 1--144, 2006.

\bibitem{gerkethesis}
S.~Gerke, ``Weighted colouring and channel assignment,'' Ph.D. dissertation,
  University of Oxford, March 2000.

\bibitem{gerke}
S.~Gerke and C.~McDiarmid, ``Graph imperfection i,'' \emph{Journal of
  Combinatorial Theory, Series B}, vol.~83, pp. 58--78, 2001.

\bibitem{gerke2}
------, ``Graph imperfection ii,'' \emph{Journal of Combinatorial Theory,
  Series B}, vol.~83, pp. 79--101, 2001.

\bibitem{fanoutsplitting}
J.~Hayes, R.~Breault, and M.~Mehmet-Ali, ``Performance analysis of a multicast
  switch,'' \emph{Communications, IEEE Transactions on}, vol.~39, no.~4, pp.
  581--587, 1991.

\bibitem{tracey}
T.~Ho, D.~R. Karger, M.~M\'edard, and R.~K\"{o}tter, ``Network coding from a
  network flow perspective,'' in \emph{Proceedings of IEEE International
  Symposium on Inform. Theory}, 2003.

\bibitem{rlc}
T.~Ho, M.~M\'{e}dard, R.~K\"{o}tter, M.~Effros, J.~Shi, and D.~R. Karger, ``A
  random linear coding approach to mutlicast,'' \emph{IEEE Transaction on
  Information Theory}, vol.~52, pp. 4413--4430, 2006.

\bibitem{harish}
T.~Ho and H.~Viswanathan, ``Dynamic algorithms for multicast with intra-session
  network coding,'' in \emph{43rd Allerton Annual Conference on Communication,
  Control and Computing}, 2005.

\bibitem{jaggi}
S.~Jaggi, P.~A. Chou, and K.~Jain, ``Low complexity algebraic multicast network
  codes,'' in \emph{Proceedings of IEEE International Symposium on Inform.
  Theory}, 2003.

\bibitem{karp}
R.~M. Karp, ``Reducibility among combinatorial problems,'' in \emph{Complexity
  of Computer Computations}, R.~E. Miller and J.~W. Thatcher, Eds.\hskip 1em
  plus 0.5em minus 0.4em\relax Plenum Press, 1972, pp. 85--103.

\bibitem{codingforspeedup}
M.~Kim, J.~K. Sundararajan, and M.~M\'{e}dard, ``Network coding for speedup in
  switches,'' in \emph{Proceedings of IEEE ISIT}, 2007.

\bibitem{comparingimperfection}
\BIBentryALTinterwordspacing
A.~M. C.~A. Koster and A.~K. Wagler, ``Comparing imperfection ratio and
  imperfection index for graph classes,'' \emph{RAIRO -- Operations Research},
  2008, to appear. [Online]. Available:
  \url{http://opus.kobv.de/zib/volltexte/2005/883/}
\BIBentrySTDinterwordspacing

\bibitem{algebraic}
R.~K\"{o}tter and M.~M\'{e}dard, ``An algebraic approach to network coding,''
  \emph{IEEE/ACM Transaction on Networking}, vol.~11, pp. 782--795, 2003.

\bibitem{mpt}
M.~Kvasnica, P.~Grieder, and M.~Baotic, ``Multi-parametric toolbox,''
  \emph{http://control.ee.ethz.ch/mpt/}, 2004.

\bibitem{LYC}
S.-Y.~R. Li, R.~W. Yeung, and N.~Cai, ``Linear network coding,'' \emph{IEEE
  Transaction on Information Theory}, vol.~49, pp. 371--381, 2003.

\bibitem{lincostello}
S.~Lin and D.~J. Costello, \emph{Error Control Coding: Fundamentals and
  Applications}.\hskip 1em plus 0.5em minus 0.4em\relax Prentice Hall, 1983.

\bibitem{marsan}
M.~A. Marsan, A.~Bianco, P.~Giaccone, E.~Leonardi, and F.~Neri, ``Multicast
  traffic in input-queued switches: Optimal scheduling and maximum
  throughput,'' vol.~11, pp. 465--477, June 2003.

\bibitem{100forunicast}
N.~McKeown, V.~Anantharam, and J.~Walrand, ``Achieving 100\% throughput in an
  input-queued switch,'' in \emph{Proceedings of IEEE INFOCOM}, 1996, pp.
  296--302.

\bibitem{conjecture}
M.~M\'{e}dard, M.~Effros, T.~Ho, and D.~R. Karger, ``On coding for
  non-mutlicast networks,'' in \emph{Proceedings of the 41st Annual Allerton
  Conference on Communication Control and Computing}, Monticello, Illinois,
  October 2003.

\bibitem{speedup}
Y.~Oie, M.~Murata, K.~Kubota, and H.~Miyahara, ``Effect of speedup in
  nonblocking packet switch,'' in \emph{Proceedings of IEEE International
  Conference on Communications}, vol.~1, 1989, pp. 410--414.

\bibitem{combinatorics}
A.~Schrijver, \emph{Combinatorial Optimization: Polyhedra and
  Efficiency}.\hskip 1em plus 0.5em minus 0.4em\relax Springer Verlag, 2003.

\bibitem{sch}
------, \emph{Combinatorial {O}ptimization: {P}olyhedra and
  {E}fficiency}.\hskip 1em plus 0.5em minus 0.4em\relax Springer Verlag, 2003.

\bibitem{jaykumar}
J.~K. Sundararajan, S.~Deb, and M.~M\'{e}dard, ``Extending the birkhoff-von
  neumann switching strategy for multicast -- on the use of optical splitting
  in switches,'' \emph{IEEE Journal on Selected Areas in Communications -
  Optical Communications and Networking Series}, 2007.

\bibitem{conflictgraphs}
J.~K. Sundararajan, M.~M\'{e}dard, R.~K\"otter, and E.~Erez, ``A systematic
  approach to network coding problems using conflict graphs,'' in
  \emph{Proceedings of the UCSD Workshop on Information Theory and its
  Applications}, San Diego, CA, February 2006.

\bibitem{ucsdita}
J.~K. Sundararajan, M.~M\'edard, R.~K\"{o}tter, and E.~Erez, ``A systematic
  approach to network coding problems using conflict graphs,'' in
  \emph{Proceedings of the UCSD Workshop on Inform. Theory and its
  Applications}, San Diego, Feb. 2006.

\bibitem{ARQforNC}
J.~K. Sundararajan, D.~Shah, and M.~M\'{e}dard, ``{ARQ} for network coding,''
  in \emph{Proceedings of IEEE ISIT}, 2008.

\bibitem{tassiulas}
L.~Tassiulas, ``Linear complexity algorithms for maximum throughput in radio
  networks and input queued switches,'' in \emph{Proceedings of IEEE INFOCOM},
  vol.~2, 1998, pp. 533--539.

\bibitem{tassephrem}
L.~Tassiulas and A.~Ephremides, ``Stability properties of constrained queueing
  systems and scheduling policies for maximum throughput in multihop radio
  networks,'' \emph{IEEE Transactions on Automatic Control}, vol.~37, no.~12,
  pp. 1936--1948, 1992.

\bibitem{neumann}
J.~von Neumann, ``A certain zero-sum two-person game equivalent to the optimal
  assignment problem,'' \emph{Contributions to the Theory of Games}, vol.~2,
  pp. 5--12, 1953.

\end{thebibliography}

\end{document}